\documentclass[fleqn,usenatbib]{mnras}

\usepackage[T1]{fontenc}
\usepackage{ae,aecompl}

\usepackage{graphicx}	
\usepackage{amsmath}	
\usepackage{amssymb}	
\usepackage{pdflscape}
\usepackage{afterpage}
\usepackage{rotating}
\usepackage{xcolor}
\usepackage{newtxtext,newtxmath}

\newcommand{\nustar}{\textit{NuSTAR}}

\newcommand{\xmm}{{\it XMM-Newton}}

\newcommand{\red}{\textcolor{black}}
\newcommand{\ergs}{erg~cm$^{-2}$~s$^{-1}$}

\newcommand{\green}{\textcolor{black}}

\newcommand{\src}{IRAS~13224}
\newcommand{\srcfull}{IRAS~13224$-$3809}
\newcommand{\second}{\textcolor{black}}

\title[Spectral Fitting of the NLS1 IRAS~13224$-$3809]{\xmm\ Observations of the Narrow-Line Seyfert 1 Galaxy \srcfull: X-ray Spectral Analysis II.}

\author[J. Jiang et al.]{
Jiachen Jiang$^{1,2}$\thanks{E-mail: jcjiang@mail.tsinghua.edu.cn}, Thomas Dauser$^{3}$,
Andrew C. Fabian$^{1}$,
William N. Alston$^{4}$, \newauthor
Luigi C. Gallo$^{5}$,
Michael L. Parker$^{1}$ and
Christopher S. Reynolds$^{1}$
\\
$^{1}$Institute of Astronomy, University of Cambridge, Madingley Road, Cambridge CB3 0HA, UK\\
$^{2}$Department of Astronomy, Tsinghua University, Beijing 100084, China\\
$^{3}$Dr Karl Remeis-Observatory and Erlangen Centre for Astroparticle Physics, Sternwartstr. 7, D-96049 Bamberg, Germany\\
$^{4}$European Space Agency, European Space Astronomy Centre, E-28691 Villanueva de la Ca\~nada, Spain\\
$^{5}$Department of Astronomy and Physics, Saint Mary's University, 923 Robie Street, Halifax, NS, B3H 3C3, Canada
}
 
\date{Accepted XXX. Received YYY; in original form ZZZ}

\pubyear{2021}

\begin{document}
\label{firstpage}
\pagerange{\pageref{firstpage}--\pageref{lastpage}}
\maketitle

\begin{abstract}
\green{Previously, we modelled the X-ray spectra of the narrow-line Seyfert 1 galaxy \srcfull\ using a disc reflection model with a fixed electron density of $10^{15}$\,cm$^{-3}$. An additional blackbody component was required to fit the soft X-ray excess below 2\,keV.} In this work, we analyse simultaneously five flux-resolved \xmm\ spectra of this source comprising data collected over 2\,Ms. A disc reflection model with an electron density of $n_{\rm e}\approx10^{20}$\,cm$^{-3}$ and an iron abundance of $Z_{\rm Fe}=3.2\pm0.5Z_{\odot}$ is used to fit \green{the broad-band spectra} of this source. \green{No additional component is required to fit the soft excess.} Our best-fit model provides consistent measurements of black hole spin and disc inclination angle as previous models where a low disc density was assumed. In the end, we calculate the average \green{illumination} distance between the corona and the reflection region in the disc of \srcfull\ based on best-fit density and ionisation parameters, which changes from \green{0.43$\sqrt{f_{\rm AD}/f_{\rm INF}}$\,$r_{\rm g}$ in the lowest flux state to 1.71$\sqrt{f_{\rm AD}/f_{\rm INF}}$\,$r_{\rm g}$ in the highest flux state assuming a black hole mass of $2\times10^{6}M_{\odot}$.  $f_{\rm AD}/f_{\rm INF}$ is the ratio between the flux of the coronal emission that reaches  the accretion disc and infinity.  This ratio depends on the geometry of the coronal region in \srcfull. So we only discuss its value based on the simple `lamp-post' model,} although detailed modelling of the disc emissivity profile of \srcfull\ is required in future to reveal the exact geometry of the corona.
\end{abstract}




\section{Introduction}

\srcfull\ (\src\ hereafter) is a narrow-line Seyfert 1 galaxy \citep[NLS1,][]{gallo18} hosting a supermassive black hole of $m_{\rm BH}=M_{\rm BH}/M_{\odot}\approx2\times10^{6}$ \citep{zhou05,alston20}. Detailed disc reflection modelling of the broad Fe K and Fe L emission lines in the X-ray band \citep{ponti10, fabian13} suggests a maximally spinning black hole in the centre \citep{fabian13, chiang15, jiang18}. \src\ is well known for its complex and rapid X-ray variability that is dominated by varying continuum emission \citep{boller97, gallo04, fabian13, buisson18, jiang18, alston19}. \red{Besides}, multiple blueshifted absorption features have been found in the X-ray spectra of \src, which may correspond to a highly variable outflow with a line-of-sight velocity of $\approx0.24c$ \citep{parker17a, parker17b, jiang18, pinto18} or a thin layer of high-ionisation matter that is co-rotating with the innermost region of the accretion disc \citep{fabian20}.

In the first paper of this series \citep{jiang18}, we analysed the spectra of \src\ that were extracted from the \xmm\ observing campaign of this source in 2016. We followed the same approach as in \citet{chiang15}: a disc reflection model with a fixed disc density of $n_{\rm e}=10^{15}$\,cm$^{-3}$ was used on an earlier and smaller dataset. A super-solar iron abundance of $Z_{\rm Fe}\approx24$\,$Z_{\odot}$ was inferred by this model. An additional blackbody component was required to model the excess emission in the soft X-ray band. This blackbody component followed a flux--temperature relation of $F \propto T^{4}$, suggesting an emission region with a constant area, e.g. the inner accretion disc \citep{chiang15}. Reverberation studies also support the disc origin of the soft excess emission in \src\ \citep{ponti10}. Detailed reverberation lag analyses based on a simple lamppost model suggest \red{an} extended coronal region of $h\approx10-20$\,$r_{\rm g}$ in the high flux states and a more compact coronal region of $h\leq10$\,$r_{\rm g}$ in the lower flux states of \src\ \citep{alston20},  similar to  measurements of micro-lensed quasars \citep{chartas17} and X-ray occultation events \citep[e.g.][]{risaliti99,gallo21}. 

\begin{figure}
    \centering
    \includegraphics{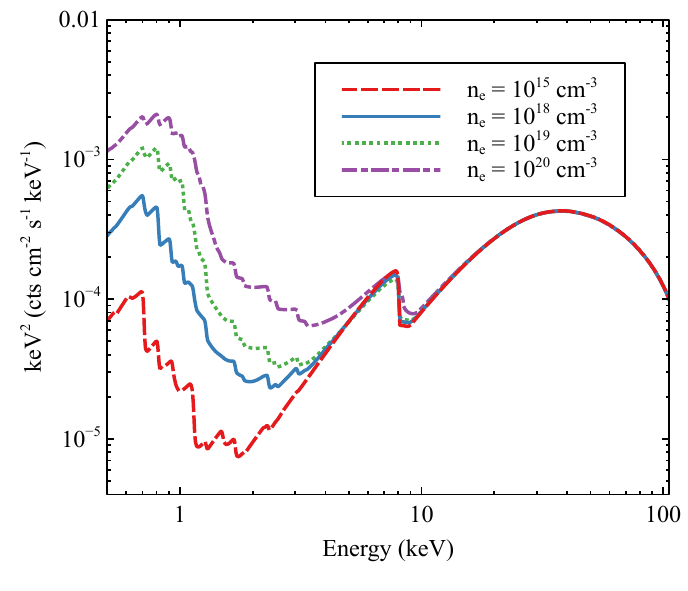}
    \caption{\green{Relativistic disc reflection spectra for $n_{\rm e}=10^{15}$\,cm$^{-3}$ (red dashed line), $10^{18}$\,cm$^{-3}$ (blue solid line), $10^{19}$\,cm$^{-3}$ (green dotted line), $10^{20}$\,cm$^{-3}$ (purple dash-dotted line). An ionisation parameter of $\xi=10$\, erg\,cm\,s$^{-1}$ and a power-law illuminating spectrum of $\Gamma=2$ are used in our calculations. A simple lamppost geometry with $h=2$\,$r_{\rm g}$ is assumed for the corona. A high BH spin of $a_{*}=0.99$ and an inclination angle of $i=65^{\circ}$ are used.}}
    \label{pic_ne}
\end{figure}

In this work, we present a reflection-based model for the \xmm\ spectra of \src. In particular, a disc model with a variable disc density parameter ($n_{\rm e}$) is considered \citep{ross07,garcia16}. The previous assumption in the disc reflection model of $n_{\rm e}=10^{15}$\,cm$^{-3}$ was only appropriate for the disc around a massive super-massive black hole both theoretically and observationally \citep[e.g. $M_{\rm BH}\geq10^{8}$,][]{shakura73, jiang18d, jiang19b}. At a higher density, free-free absorption become important up to X-ray energies, and the reflection spectrum of the disc produces a blackbody-like feature in the soft X-ray band \citep[][\green{or see Fig.\ref{pic_ne}}]{garcia16}. Such a feature can explain the soft excess emission commonly seen in the X-ray spectra of unobscured Seyfert 1 galaxies \citep[][]{mallick18,garcia18,jiang19b,jiang20b}. Moreover, the disc density parameter inferred by reflection spectroscopy also shows a correlation with the mass accretion rate of the accretion disc according to the studies of stellar-mass black holes \citep{jiang19, jiang20}. Last but not least, a high density disc reflection model may be able to solve the problem of super-solar iron abundances obtained by previous models \citep{tomsick18,jiang19b}. Lower inferred iron abundances are due to the model requiring a higher contribution of disc reflection to the total X-ray spectrum \citep{jiang19b}. 

In Section\,\ref{data}, we introduce our high density disc model for the \xmm\ spectra of \src. In Section\,\ref{calculations}, we estimate the size of the coronal region for different flux states based on the best-fit parameters given by our model. In Section\,\ref{conclusions}, we discuss and conclude our results. 

\section{Data Analysis} \label{data}

In this section, we introduce a high density disc-based model to explain the spectral variability of \src. We first present an overview of the broad band spectra of \src\ in five different flux states. Then we focus on the spectral modelling using high density disc reflection model.

\subsection{Flux-Resolved Spectra}

\begin{table}
    \caption{\red{A list of \xmm\ observations of \src\ used in this work.}}
    \label{tab_obs}
    \centering
    \begin{tabular}{ccc}
	\hline\hline				
Obs ID	&	Time	&	Duration (s)	\\
\hline
110890101	&	2002-01-19 02:35:23	&	64019	\\
673580101	&	2011-07-19 02:13:11	&	133039	\\
673580201	&	2011-07-21 02:17:09	&	132443	\\
673580301	&	2011-07-25 01:49:41	&	129438	\\
673580401	&	2011-07-29 01:27:48	&	134736	\\
780560101	&	2016-07-08 19:33:33	&	141300	\\
780561301	&	2016-07-10 19:25:31	&	141000	\\
780561401	&	2016-07-12 19:17:06	&	138100	\\
780561501	&	2016-07-20 18:44:57	&	140800	\\
780561601	&	2016-07-22 18:36:58	&	140800	\\
780561701	&	2016-07-24 18:28:28	&	140800	\\
792180101	&	2016-07-26 18:18:44	&	141000	\\
792180201	&	2016-07-30 18:02:21	&	140500	\\
792180301	&	2016-08-01 17:54:51	&	140500	\\
792180401	&	2016-08-03 17:47:25	&	140800	\\
792180501	&	2016-08-07 17:40:58	&	138000	\\
792180601	&	2016-08-09 18:29:52	&	136000	\\
\hline\hline
    \end{tabular}

\end{table}

We consider all the archival \xmm\ observations of \src\ in this work, including the 1.5\,Ms observing campaign in 2016 (PI: Fabian, A. C.) and previous observations with a length of around 600\,ks. A full list of these observations can be found in \red{Table\,\ref{tab_obs}}. We focus on the broad band spectral variability shown by the EPIC-pn (pn hereafter) observations. All the pn observations were operated in the Large Window mode. SAS v.1.3 was used to extract spectra from filtered clean event lists. We follow the standard data reduction method as already introduced in the first paper of this series \citep{jiang18}. The same source regions as in the first paper of this series \citep{jiang18} were used after using the EPATPLOT tool to check pile-up effects. 

\src\ shows rapid variability in the X-ray band \citep{boller03,fabian13,chiang15}. In order to better understand the spectral variability of this source, we divided all the observations into five different flux intervals. 
Five intervals share a similar number of total photon counts. The TABGTIGEN tool was used for this purpose. This method was also used in our previous analysis of the data \citep{parker17a,jiang18}. The final spectra respectively have an average observed flux of $\log(F/$erg cm$^{-2}$ s$^{-1})=-11.948\pm0.003$ (F1, \red{<1.1 cts\,s$^{-1}$}), $-11.614\pm0.003$ (F2, \red{1.1--1.7 cts\,s$^{-1}$}), $-11.454\pm0.003$ (F3, \red{1.7--2.4 cts\,s$^{-1}$}), $-11.313\pm0.002$ (F4, \red{2.4--3.4 cts\,s$^{-1}$}) and $-11.108\pm0.002$ (F5, \red{>3.4 cts\,s$^{-1}$}) in the 0.5--10\,keV band.

The spectra are grouped to have a minimum number of 20 counts per bin, and no group is narrower than 1/3 of the full width half maximum resolution of the pn instrument. XSPEC v.12.11.01\citep{arnaud85} is used to analyse the spectra.

\begin{figure}
    \centering
    \includegraphics{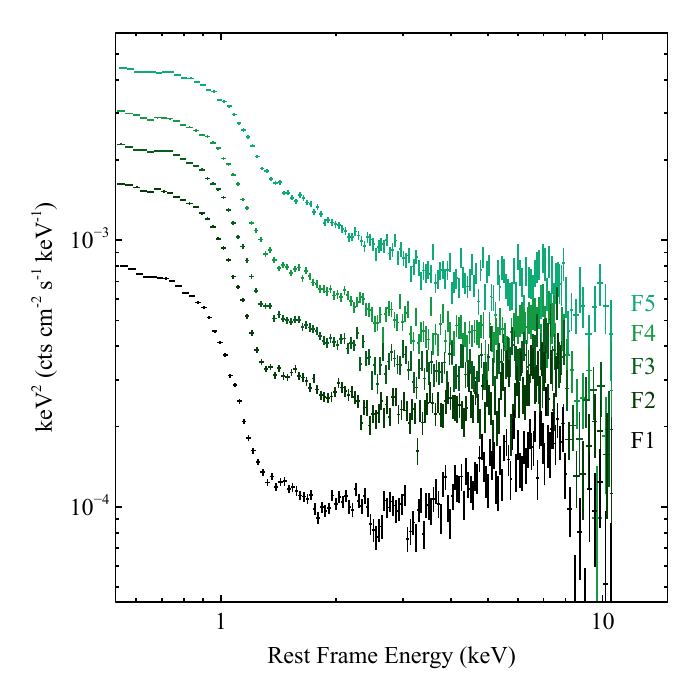}
    \caption{Five flux-resolved pn spectra of \src\ extracted from all the \xmm\ observations in the archive. The spectra are unfolded using a model that is constant across the energy band for demonstration purposes only.}
    \label{pic_eeuf}
\end{figure}

Fig. \ref{pic_eeuf} shows these five flux-resolved spectra of \src. They are unfolded using a constant model to remove the effect of instrumental response only for demonstration purposes. As shown in Fig.\,\ref{pic_eeuf}, different energy bands show flux changes by a different factor between two states. For instance, F5 has a flux approximately 9 times higher than F1 in the 2--4\,keV band. A lower flux increase factor is found in the other bands, e.g. a factor of 4 between 4--10\,keV band a factor of 7 between 2--4\,keV. Similar conclusions are found in \citet{alston20} where RMS spectra are analysed. Previous reflection modelling suggests that the 2--4\,keV band of \src\ is dominated by highly variable coronal emission while the 4--10\,keV band is dominated by less variable reflection from the innermost region of the disc \citep{fabian09,chiang15,jiang18,alston20}. A higher fraction of disc reflection is found in low flux states than in high flux states \citep{jiang18}.

\subsection{Model Set-Up} \label{mo_set}

In this section, we introduce a high-density disc model for the reflection spectra in the five spectra of \src\ in different flux states\footnote{\red{We show a low-density disc model for the reflection spectrum of \src\ in Appendix\,\ref{low_ne}, where an electron density of $10^{15}$\,cm$^{-3}$ is assumed for the disc.}}.

The coronal emission of \src\ is modelled using the thermal Comptonisation model \texttt{nthcomp} \citep{zycki99}. We consider a disc-blackbody spectrum for the seed photons, which is characterised by the inner disc temperature $kT_{\rm db}$. \red{This parameter determines the low energy rollover of a thermal Comptonisation spectrum and cannot be constrained by X-ray spectra alone due to Galactic absorption.}  We therefore fix this parameter at 10\,eV, the typical value for an accretion disc around a supermassive black hole.  The electron temperature of the corona is fixed at 100\,keV\footnote{\red{In Appendix \ref{kt}, we show the model for the soft X-ray data is consistent when a low $kT_{\rm e}=10$\,keV is used, except for a slightly harder photon index.}}.

The rest-frame disc reflection spectrum is calculated using the \texttt{reflionx} code\footnote{Another popular model \texttt{xillverd} can be used to calculate rest-frame \red{disc} reflection spectra too \citep{garcia10}.  \red{In Appendix\,\ref{relxilld}, we present a fit for the reflection spectrum of \src\ using an \texttt{xillverd}-based model (\texttt{relxilld}). We find that the fit is unsatisfactory due to its limited range of $n_{\rm e}$. Therefore,} we choose \texttt{reflionx} over the public version of \texttt{xillverd} for a wider range of the disc density parameter in \texttt{reflionx}. } \citep{ross07}. The \texttt{nthcomp} model is used as the illuminating spectrum \citep{jiang20b}. $kT_{\rm db}$, $kT_{\rm e}$ and photon index $\Gamma$ of \texttt{reflionx} are all linked to the corresponding parameters in \texttt{nthcomp}. The other free parameters are the ionisation state ($\xi$), the iron abundance ($Z_{\rm Fe}$) and the electron density within the Thomson depth of the disc $n_{\rm e}$. $Z_{\rm e}$ is calculated using the solar abundances calculated in \citet{morrison83}.

The convolution model \texttt{relconv} is applied to the rest frame reflection model \texttt{reflionx} to account for relativistic effects \citep{dauser10}. A broken power-law emissivity profile is considered. The spin of the central black hole, the iron abundance and the inclination angle of the disc are not expected to change on observable timescales. We therefore link these parameters for all five spectra.

Previous studies by \citet{parker17a,jiang18,pinto18} found a highly variable blueshifted ionised absorption line features, suggesting evidence for an ultra-fast outflow with a line-of-sight velocity of $\approx0.24c$. We model the absorption features using the photon ionisation plasma model \texttt{xstar} \citet{kallman01}. \red{The \texttt{xstar} grid in this work is calculated using a power law-shaped illuminating spectrum with $\Gamma=2$. Free parameters include the column density ($N^{\prime}_{\rm H}$), the ionisation\footnote{The prime symbol is to distinguish the ionisation parameter of the wind from the same parameter of the \red{disc}.} ($\xi^{\prime}$), the iron abundance ($Z^{\prime}_{\rm Fe}$) and the redshift parameter ($z$).} Note that solar abundance in the \texttt{xstar} model is calculated by \citet{grevesse96}, which is different from the values used by \texttt{reflionx}. Therefore, the iron abundance parameters of these components are not linked in our model.

In addition, the \texttt{tbnew} model is used to account for Galactic absorption \citep{wilms00}. The \texttt{cflux} model is used to calculate the unabsorbed flux of each component in the full 0.5--10\,keV band. The total model is \texttt{tbnew*xstar* (cflux*relconv*reflionx + cflux*nthcomp)} in the XSPEC format.

\begin{figure*}
    \centering
    \includegraphics[width=18cm]{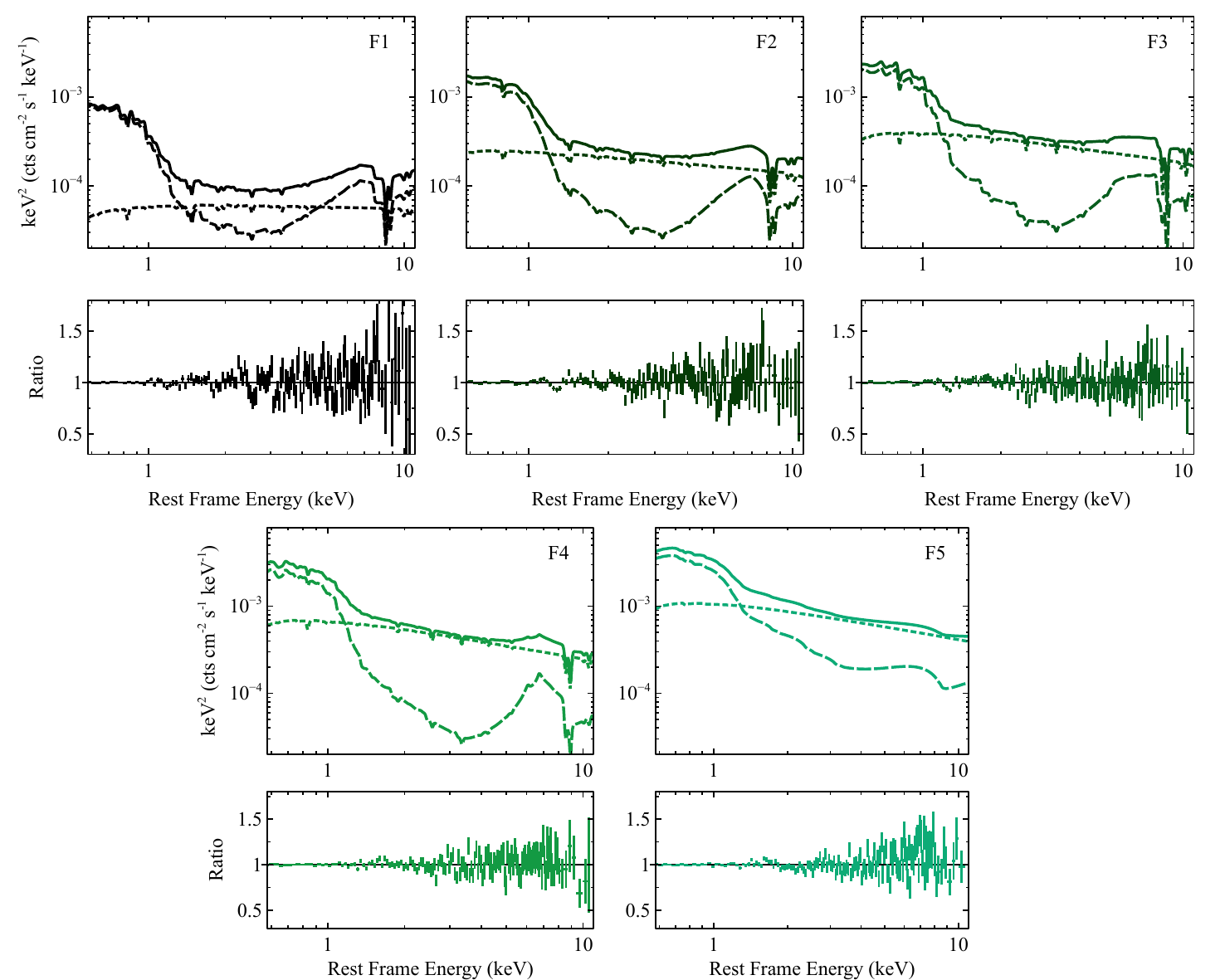}
    \caption{Top panels: best-fit models. Solid lines: total model; dashed lines: disc reflection; dotted line: coronal emission. Bottom panels: corresponding data/model ratio plots.}
    \label{pic_fit}
\end{figure*}

\subsection{Results}

The model introduced above is able to describe the broad band spectra of \src\ very well with a total $\chi^{2}/\nu=864.32/798$. The best-fit models and corresponding data/model ratio plots are shown in Fig.\,\ref{pic_fit}. The best-fit model parameters can be found in Table \ref{tab_fit}. \green{Note that the high density models do not require any additional component for the soft excess flux seen below 2 keV.} 

Our best-fit absorption model suggests an outflow with a line-of-sight velocity of $0.21-0.24c$. The ionisation state of the outflow during F1 and F2 are significantly lower than that during F3 and F4. The highest flux state F5 shows no significant evidence for blueshifted absorption features. We therefore fix the ionisation parameter $\xi^{\prime}$ and the redshift parameter $z$ of \texttt{xstar} for F5 at the best-fit values for F4. Only an upper limit of the column density of the outflow is found (a 90\% confidence range of $N^{\prime}_{\rm H}<2\times10^{22}$\,cm$^{-2}$). Similar conclusions were found in previous analyses \citep{parker17a, jiang18, pinto18}. Interested readers may refer to \citet{parker17a} for more discussion about the rapid variability of the absorption features. 

The best-fit \texttt{relconv} model indicates a highly spinning black hole in \src\ (>0.98). The inclination of the disc is estimated to be within a 90\% confidence range of 64--69$^{\circ}$. They are consistent with previous measurements when a fixed disc density of $n_{\rm e}=10^{15}$\,cm$^{-3}$ is assumed \citep{gallo04,fabian13,chiang15,jiang18}. 

\begin{figure}
    \centering
    \includegraphics{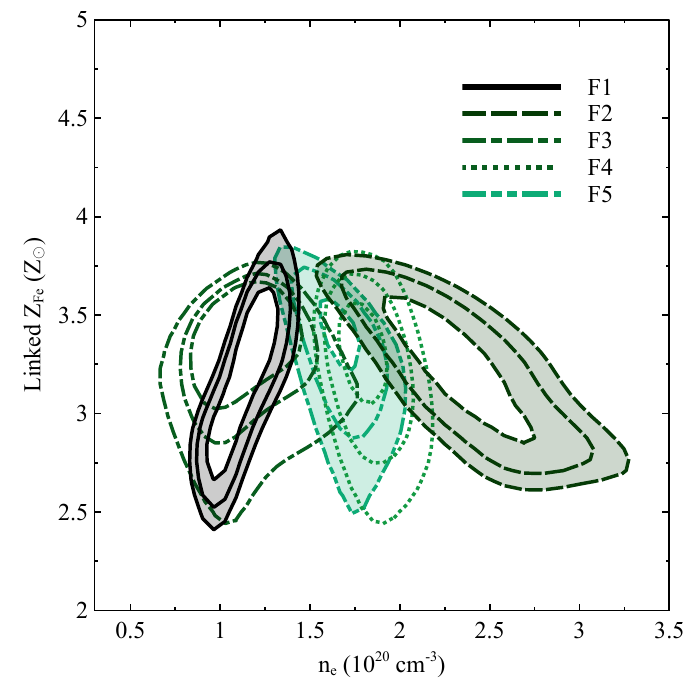}
    \caption{\green{$\chi^{2}$ distributions on the $Z_{\rm Fe}$ and $n_{\rm e}$ parameter plane. $Z_{\rm Fe}$ is linked between observations as the abundances of the disc are not expected to change on observable timescales. The contours represent 1, 2 and 3--$\sigma$ uncertainty ranges. F1: solid contours; F2: dashed contours; F3: dash-dotted contours; F4: dotted contours; F5: dash-dot-dotted contours.}}  
    \label{pic_con}
\end{figure}

We also find tentative evidence of a variable disc density parameter: during F1 and F3, the best-fit density parameter is lower than those for other flux states (F2, F4 and F5), although the density parameters for F2-F5 are consistent within their 90\% confidence ranges. \green{The uncertainty ranges of their density parameters and the iron abundance parameter are shown in Fig.\,\ref{pic_con}. When applying a high density disc model to the data of \src, one can find significantly lower disc iron abundances of $\approx3Z_{\odot}$ than previous results \citep[e.g.][]{chiang15,jiang18}.}

\green{Previously, we found the density of the inner disc in \src\ has to be higher than $10^{19}$\,cm$^{-3}$ \citep{jiang18}. In this paper, we are able to confirm the results and obtain a high disc density of around $10^{20}$\,cm$^{-3}$ for \src. Such a disc density is expected for a low-$M_{\rm BH}$ NLS1 like \src\ according to the thin disc-corona model \citep{svensson94}. Using Eq.\,8 in \citet{svensson94}, one can find a disc density of around $10^{20}$\,cm$^{-3}$ at $r=2-10$\,$r_{\rm g}$ for a $10^{6}M_{\odot}$ BH with $\dot{M}/\dot{M_{\rm Edd}}=1$ and $f_{\rm corona}=0.9$. Similar comparison between density measurements for Sy1s and the prediction of the thin disc model can be found in \citet{jiang19b}. The parameter $f_{\rm corona}$ is the fraction of the disc energy that is radiated away in the form of non-thermal emission, including disc reflection and coronal emission. The value of $f_{\rm corona}$ is estimated to be close to 1 to explain the typical spectral energy distribution of Sy1s \citep[e.g.][]{haardt91,jiang20b}.}

\begin{figure*}
    \centering
    \includegraphics[width=18cm]{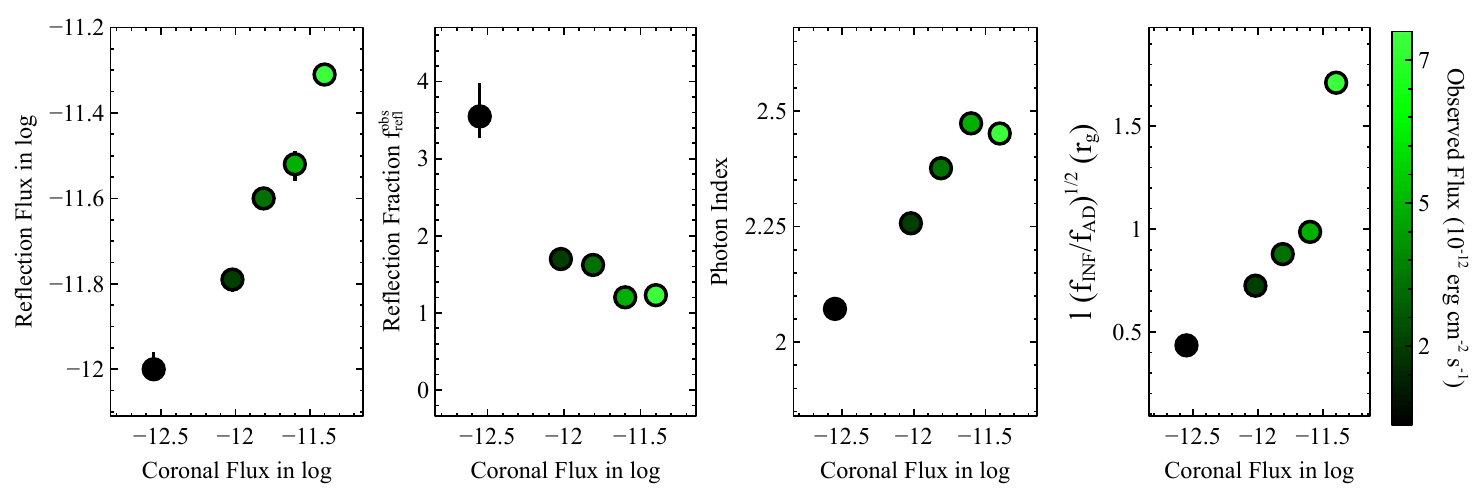}
    \caption{Best-fit parameters vs. the 0.5--10\,keV flux of the coronal emission. The flux values are in units of erg\,cm$^{-2}$\,s$^{-1}$. The colours of the points represent observed 0.5--10\,keV flux. The error bars in the third panel are smaller than the size of the points. $l\sqrt{f_{\rm INF}/f_{\rm AD}}$ in the fourth panel is estimated using the best-fit $n_{\rm e}$, $\xi$ and the 0.01--100\,keV band flux of the coronal emission. A black hole mass of $2\times10^{6}$\,$M_{\odot}$ is assumed.}
    \label{pic_paras}
\end{figure*}

According to our best-fit models, the flux of the disc reflection component ($F_{\rm refl}$) and the coronal emission ($F_{\rm pl}$) increases with the X-ray luminosity as shown in the first panel of Fig.\,\ref{pic_paras}. $F_{\rm refl}$ increases from $10^{-12}$\,\ergs\ for F1 to $5\times10^{-12}$\,\ergs\ for F5 while $F_{\rm pl}$ increases from $3\times10^{-13}$\ergs\ for F1 to  $4\times10^{-12}$\,\ergs\ for F5. 


The second panel of Fig.\,\ref{pic_paras} shows how the strength of the reflection component relative to the power-law continuum (reflection fraction) changes with X-ray flux. The reflection fraction in the figure is defined as the flux ratio between two components in the 0.5--10\,keV band. A significantly higher fraction of reflection is shown in the lower flux states. A decreasing reflection fraction along with an increasing X-ray luminosity was also found in previous analyses of \src\ when a low density disc model was used and an additional blackbody model was required to model the soft excess \citep{jiang18}.

\begin{figure}
    \centering
    \includegraphics{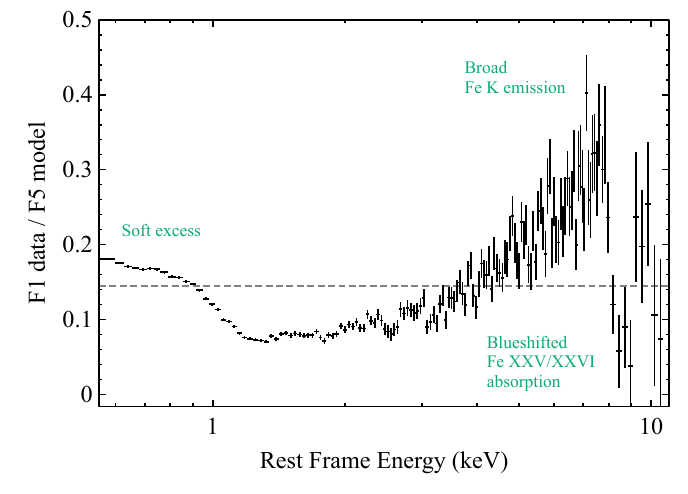}
    \caption{F1 data / F5 best-fit model ratio plot. The horizontal dashed line shows the 0.5--10\,keV observed flux ratio between F1 and F5. This ratio plot shows spectral difference between F1 and F5, including flux-dependent blueshifted Fe \textsc{xxv} and Fe \textsc{xxvi} absorption, soft excess emission and a broad Fe K emission line.}
    \label{pic_diff}
\end{figure}

In order to better demonstrate the spectral difference between the high and low flux states, we show a F1 data / F5 best-fit model ratio plot in Fig.\,\ref{pic_diff}. The grey dashed horizontal line shows the 0.5--10\,keV band flux ratio of these two states. 

The negative relation between the direct X-ray coronal emission and the reflection component can be explained by the strong light bending of the coronal emission when the corona is closer to the black hole: the flux of the continuum emission decreases as more coronal photons are lost to the event horizon and due to the large redshift closer to the black hole. The reflection spectrum of the disc is less affected by the light-bending effects \citep[e.g.][]{miniutti03, dauser13, wilkins21}. Reverberation studies of \src\ also support this conclusion \citep{alston20}. \second{{Other explanations may involve different geometries for the corona \citep[e.g. extended sizes or outflowing geometries, ][]{dauser13,steiner17,szanecki20}. Returning radiation of the inner disc \citep{riaz21} and small changes in the inner disc radius may introduce changes in the observed reflection fraction too. In the next section, we present calculations of the distance between the corona and the inner disc suggested by our best-fit spectral models. In particular, we consider the best-understood lamppost model because of the difficulty of calculating other complex geometries. Future efforts based on other theories are needed for the analysis of the X-ray data of \src.}}

\begin{table*}
    \centering
    \begin{tabular}{cccccccccccccccccccccccc}
    \hline\hline
    Model & Parameter & Unit & F1 & F2 & F3 & F4 & F5 \\
     \hline
     \texttt{tbnew} & $N_{\rm H}$ & $10^{20}$\,cm$^{-2}$ & & & $5.8\pm0.2$  & &\\
     \hline
     \texttt{xstar} & $N^{\prime}_{\rm H}$ & $10^{23}$\,cm$^{-2}$ & $3.2\pm0.2$ &  $1.1^{+0.4}_{-0.2}$ & $2.5^{+0.7}_{-0.8}$ & $4.6^{+0.5}_{-0.4}$ &  $<0.2$ & \\
     & $\log(\xi^{\prime})$ & erg cm s$^{-1}$ & $3.35^{+0.02}_{-0.03}$ & $3.39^{+0.02}_{-0.08}$ & $3.60^{+0.04}_{-0.05}$ & $3.66^{+0.03}_{-0.02}$ & 3.66 \\
     & $Z^{\prime}_{\rm Fe}$ & $Z_{\odot}$ &  & & $2.7^{+1.8}_{-0.3}$ &  & \\
     & $z$ & - & $-0.159^{+0.004}_{-0.006}$ & $-0.141^{+0.003}_{-0.004}$ & $-0.144^{+0.004}_{-0.005}$ & $-0.165^{+0.002}_{-0.003}$ & $-0.165$ \\
     \hline
     \texttt{relconv} & q1 & - & $6^{+2}_{-3}$ & $3^{+2}_{-1}$ & $7\pm2$ & $2.4^{+1.2}_{-0.3}$  & $1.7^{+0.9}_{-0.6}$ \\
      &q2 & - & $4\pm2$ & $7^{+2}_{-3}$ & $3^{+2}_{-1}$ & $2.2^{+0.8}_{-0.7}$  & $2\pm2$\\
      &$R_{\rm b}$ & $r_{\rm g}$ & $4\pm2$ & $<3$ &  $<4$ & $<20$ & $3.2^{+1.3}_{-0.6}$ \\
      &$a_*$ & - &  & & >0.98  & &\\
      &$i$ & - &  & &$67^{+2}_{-3}$  & &\\
      \hline
     \texttt{reflionx} & $n_{\rm e}$ & 10$^{20}$ cm$^{-3}$ & $0.9\pm0.2$ & $1.9^{+1.2}_{-0.3}$ & $1.3\pm0.3$ & $1.9\pm0.3$ & $1.5^{+0.3}_{-0.2}$ \\
     & $\xi$ & erg cm s$^{-1}$ & $6.5\pm0.5$ & $5\pm2$ & $10\pm2$ & $11\pm2$ & $7^{+1.3}_{-2.0}$ \\
     & $Z_{\rm Fe}$ & $Z_{\odot}$  & && $3.2\pm0.5$  & &\\
     & $\Gamma$ & - & $2.072^{+0.005}_{-0.004}$ &  $2.257^{+0.008}_{-0.006}$ & $2.376^{+0.003}_{-0.006}$ & $2.473^{+0.008}_{-0.010}$ & $2.451^{+0.015}_{-0.011}$  \\
     & $\log(F_{\rm refl})$ & erg\,cm$^{-2}$\,s$^{-1}$ & $-12.00^{+0.04}_{-0.03}$ & $-11.79^{+0.02}_{-0.03}$ & $-11.598^{+0.008}_{-0.010}$ & $-11.52^{+0.03}_{-0.04}$ & $-11.310^{+0.005}_{-0.006}$ \\
     \hline
     \texttt{nthcomp} & $\log(F_{\rm pl})$ & erg\,cm$^{-2}$\,s$^{-1}$ & $-12.55\pm0.03$ & $-12.02\pm0.02$ & $-11.81\pm0.02$ & $-11.604^{+0.008}_{-0.007}$ & $-11.400^{+0.003}_{-0.002}$ \\
     \hline
     & $f^{\rm obs}_{\rm refl}$ & - & $3.5^{+0.4}_{-0.3}$ & $1.70^{+0.11}_{-0.14}$ & $1.62^{+0.09}_{-0.08}$ & $1.20^{+0.09}_{-0.11}$ & $1.230^{+0.017}_{-0.018}$ \\
     & $\sqrt{f_{\rm INF}/f_{\rm AD}} l$ & $r_{\rm g}$ & 0.43 & 0.72 & 0.88 & 0.99 & 1.71 \\
     & $\chi^{2}/\nu$ & -&  & & 864.32/798 & & \\
    \hline\hline
    \end{tabular}
    \caption{The best-fit parameters for five flux state spectra. $f^{\rm obs}_{\rm refl}$ is defined as $F_{\rm refl}/F_{\rm pl}$ in the 0.5--10\,keV band. $\sqrt{f_{\rm INF}/f_{\rm AD}} l$ is calculated in the 0.01--100\,keV band as the ionisation parameter $\xi$ in \texttt{reflionx}. See text for more details.}
    \label{tab_fit}
\end{table*}

\section{Disc-Corona Distance} \label{calculations}

In this section, we calculate the average illumination distance $\ell$ between the corona and the illuminated disc by using the best-fit parameters obtained in Section\,\ref{data}. The estimation of $l$ is also determined by the geometry of the coronal region. In particular, we consider a lamppost geometry \red{for simplicity}, where a point-like isotropic source above the accretion disc is assumed \citep{matt91}, to provide a consistent interpretation of our energy spectra as our reverberation analyses for the same set of data in \citet{alston20}.

\subsection{The $n_{\rm e}$ and $\xi$ Parameters of the Inner Disc} \label{nemeasure}

We first calculate the average illumination distance between the corona and the disc by using the best-fit ionisation ($\xi$) and density ($n_{\rm e}$) parameters.

In the \texttt{reflionx} model, $\xi$ is defined as following by given the flux of the illuminating spectrum ($F_{\rm AD}$) and the number density of the electrons ($n_{\rm e}$),

\begin{equation}
     \xi = 4\pi \frac{F_{\rm AD}}{n_{\rm e}},
\end{equation}
where $F_{\rm AD}$ calculated between 0.01--100\,keV \citep{ross99}. $F_{\rm AD}$ can be calculated using
\begin{equation}
    F_{\rm AD} = f_{\rm AD} \frac{L_{\rm corona}}{l^{2}},
\end{equation}
where $f_{\rm AD}$ is the fraction of the coronal emission that reaches the accretion disc divided by solid angle, $L_{\rm pl}$ is the luminosity of the corona, $l$ is the average illumination distance between the disc and the corona.

Similarly, the observed coronal flux is
\begin{equation}
     F_{\rm corona} = f_{\rm INF} \frac{L_{\rm corona}}{D^{2}},
\end{equation}
where $f_{\rm INF}$ is the fraction of the coronal emission that reaches infinity divided by solid angle, and $D$ is the distance of the source.

By combining the equations above, we obtain the average illumination distance between the corona and the disc is
\begin{equation} \label{eq:fboostMeasured}
     \frac{l}{f} = \sqrt{4\pi\frac{F_{\rm corona}}{n_{\rm e}\xi}} D ,
\end{equation} where $f=\sqrt{\frac{f_{\rm AD}}{f_{\rm INF}}}$.

In our work, the absorption-corrected coronal flux $F_{\rm corona}$ is calculated in the same energy band (0.01--100\,keV) as the ionisation parameter. 

The values of $l/f$ are calculated by using Eq.\,\ref{eq:fboostMeasured} for all the flux states using their best-fit parameters, and the results are in Table\,\ref{tab_fit} and the last panel of Fig.\,\ref{pic_paras}.  A luminosity distance of $288$\,Mpc reported on the NED website is used\footnote{We assume $H_{0}=73.00$\,km\,sec$^{-1}$\,Mpc$^{-1}$, $\Omega_{\rm matter}=0.27$ and $\Omega_{\rm vacuum}=0.73$.} \citep{jones09}. 

\red{The lowest value of $l\approx0.43f\times\frac{2\times10^{6}}{m_{\rm BH}}$\,$r_{\rm g}=8.6\times10^{5}\frac{f}{m_{\rm BH}}$\,$r_{\rm g}$ is found for the lowest X-ray flux state (F1). The highest flux state (F5) has the highest value of $l\approx1.71f\times\frac{2\times10^{6}}{m_{\rm BH}}$\,$r_{\rm g}=3.4\times10^{6}\frac{f}{m_{\rm BH}}$\,$r_{\rm g}$.} 


\subsection{An Interpretation Based on the  Lamppost Model} \label{calculations2}

So far, we have calculated the average illumination distance between the corona and the disc $l$ using the ratio $f=\sqrt{f_\mathrm{AD}/f_\mathrm{INF}}$. The exact values of this ratio depends on the geometry of the coronal region and the spacetime around the BH. In this work, we consider an isotropic point-like corona, namely the lamppost model \citep{matt91}, for simplicity and consistency with our work in \citet{alston20}.

\subsubsection{Calculations of $f_\mathrm{AD}/f_\mathrm{INF}$}

In order to calculate $f_\mathrm{AD}/f_\mathrm{INF}$, the flux incident on the accretion disc has to be calculated. {Assuming an isotropic emission of the corona, this ratio would be unity in flat space time.  Due to the strong relativistic effects close to the black hole, however, several additional factors have to be taken into account.}

{First of all, light bending leads to an increasing fraction of photons bent towards the accretion disc, which is parametrized by the reflection fraction $f_\mathrm{refl}$ \citep[see][]{dauser14} and can easily reach values of 10 and larger for a compact corona\footnote{The reflection fraction here is different from $f^{\rm obs}_{\rm refl}$ shown in Table\,\ref{tab_fit}. $f^{\rm obs}_{\rm refl}$ is defined as the flux ratio between the reflection component and the thermal Comptonisation component in the 0.5--10\,keV band.}. Additionally, the energy shifts from the corona to the accretion disc $g_\mathrm{AD}$ and from the corona directly to the observer $g_\mathrm{INF}$ lead to a change in flux. The exact amount of change in flux depends on the spectral shape of the corona, which we parametrise by the photon index $\Gamma$. Using the energy shifts from the lamp post geometry  \citep[see, e.g., ][]{dauser13}, the ratio can be expressed as}

\begin{equation}
    \frac{f_\mathrm{AD}}{f_\mathrm{INF}} = 
    f_\mathrm{refl} \left(\frac{g_\mathrm{AD}}{g_\mathrm{INF}}\right)^\Gamma
\end{equation}

\begin{figure}
    \centering
    \includegraphics[width=\columnwidth]{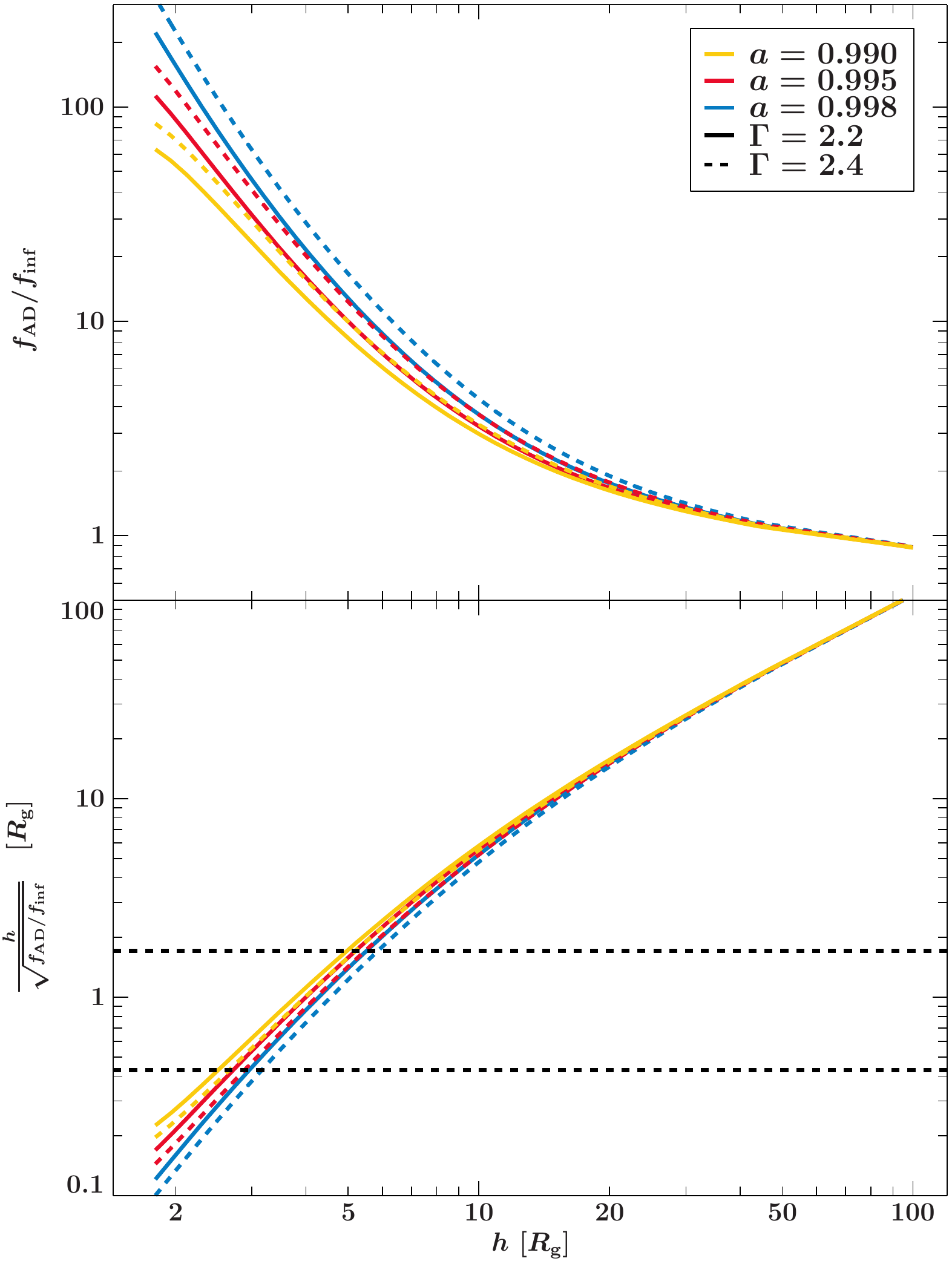}
    \caption{Upper panel: Ratio between flux incident on the accretion disc to the directly observed coronal flux for a range of parameters as determined for IRAS~13224. Lower panel: luminosity distance using the flux ratio from the upper panel. Dashed lines are the lowest and largest values as determined from the data. }
    \label{fig:fboost}
\end{figure}

{Figure~\ref{fig:fboost}a shows the flux ratio $f_\mathrm{AD}/f_\mathrm{inf}$ for the range of parameters allowed within the rage of our best fits. For very low heights of the corona, the flux incident on the accretion disc can be more than 100 times larger than the observed flux from the corona. In order to be able to compare our best fit parameters with these calculations, Fig.~\ref{fig:fboost}b plots $d_\mathrm{lum}(h) = h/\sqrt{f_\mathrm{AD}/f_\mathrm{INF}}$.}

\subsubsection{The size of the reflection region in the disc} \label{size_cal}

\begin{figure}
    \centering
    \includegraphics[width=\columnwidth]{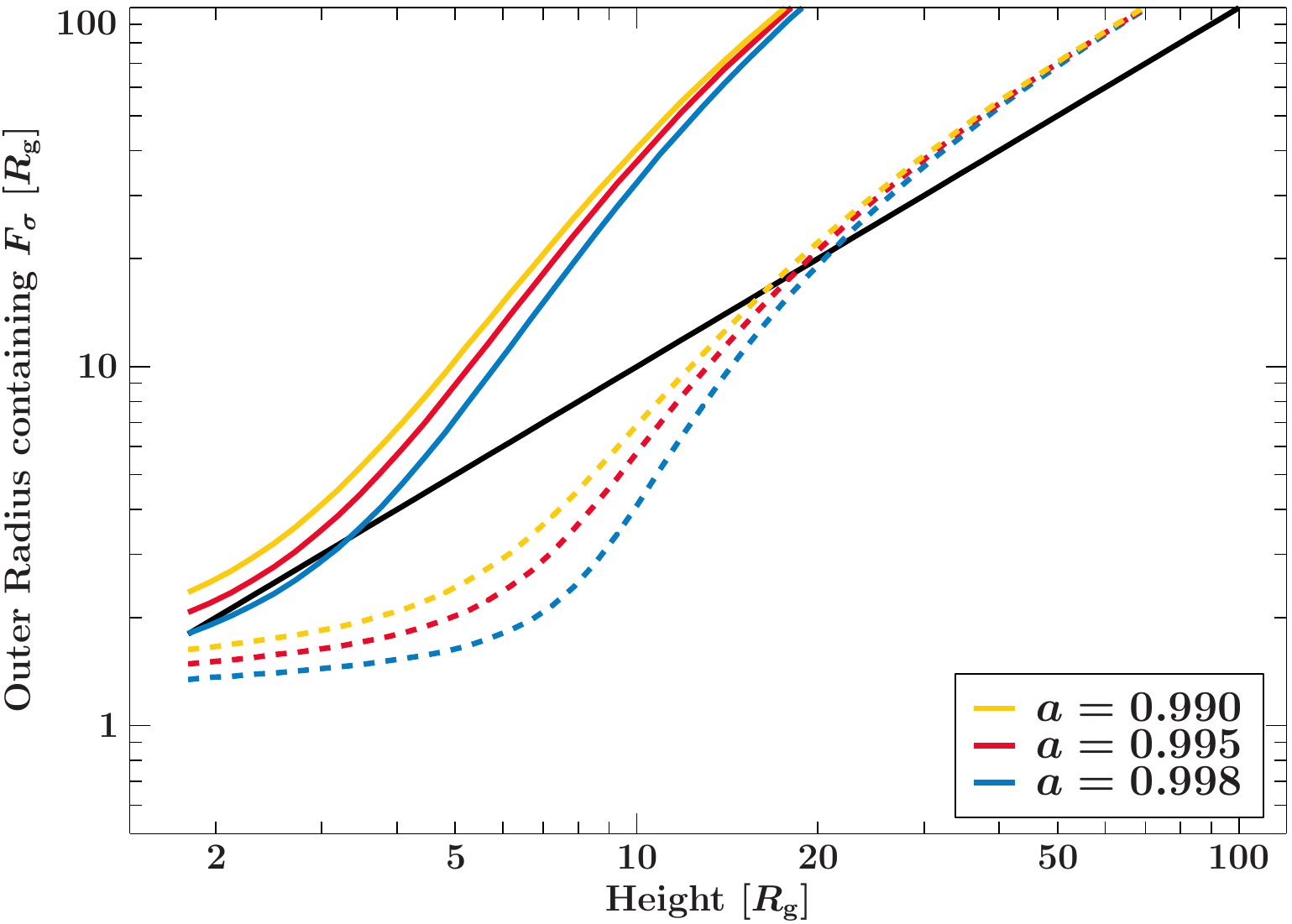}
    \caption{The outer radius, which encloses 90\% (solid) and 50\% (dashed) of the total flux incident on the accretion disc. To guide the eye, the solid line is plotted for $r=h$.} 
    \label{fig:radiusEnclosingFlux}
\end{figure}

Secondly, we estimate the size of the reflection region in the disc, based on which we argue that the average distance between the corona and the reflection region ($l$) shares a similar value with the height of the corona ($h$) when $h$ is sufficiently small. 

In Fig.~\ref{fig:radiusEnclosingFlux}, we show the outer radius defining the area which sees most of the incident flux of the corona using the same codes as in \cite{dauser13}. For instance, using 50\% of the enclosed flux as the averaged value of incident radiation, the corona can be seen that up to $h=$5\,$r_\mathrm{g}$ most of the radiation hits the disc between the ISCO and 2\,$r_\mathrm{g}$ (see the dashed lines). The average distance between the corona and the reflection region is approximately $l=\sqrt{5^2+2^2}=5.38$\,$r_\mathrm{g}$=1.08$h\approx h$. In comparison, the difference between $l$ and $h$ becomes important when the corona is very far from the disc, e.g. $l=\sqrt{r^{2}+h^{2}}>\sqrt{2}h$ when $h>20$\,$r_{\rm g}$.

{Our best-fit spectral model suggest a compact region in the inner disc where reflection happens. Therefore, $h\sqrt{f_{\rm INF}/f_{\rm AD}}$ can be approximated as $l\sqrt{f_{\rm INF}/f_{\rm AD}}$. In Fig.\,\ref{fig:fboost}, we show the values of $l\sqrt{f_{\rm INF}/f_{\rm AD}}$ for the lowest and the highest flux states of \src\ that are calculated in Section\,\ref{nemeasure}. They correspond to a height of the corona that is varying between 3-6\,$r_\mathrm{g}$. } \green{Similar calculations have also been used to study the geometry of the innermost accretion region in Cyg~X-1 \citep{zdziarski20}. Further discussion of the lamppost model can be found in \citet{niedzwiecki19} where some minor model differences are discussed.}

\section{Discussion} \label{conclusions}

\red{\subsection{Spectral Models for \src}}

Previously, the X-ray spectra of \src\ were \red{fitted} by reflection models with super-solar iron abundances of $Z_{\rm Fe}\approx20Z_{\odot}$ and a low disc density of $10^{15}$\,cm$^{-3}$ \citep[e.g.][]{chiang15,jiang18}. An additional blackbody model and two relativistic disc reflection components were required for the stacked spectrum of \src\ \citep{fabian13,jiang18}. \red{This double-reflection model was required because an averaged spectrum extracted from a wide range of flux states was considered\footnote{\red{Appendix\,\ref{stack} shows an example where a double-reflection model  similar to the one in \citet{jiang18} is used to fit the stacked spectrum of \src.}}.}

In this work, we consider all the archival \xmm\ observations of \src\ and divide the data into 5 flux-resolved intervals. We find that only one high density disc reflection model with $n_{\rm e}\approx10^{20}$\,cm$^{-3}$ and $Z_{\rm Fe}\approx3Z_{\odot}$ together with a thermal Comptonisation model is needed for each flux state. \red{The decrease of inferred iron abundances in a high density model is due to an increase in the relative strength of the reflection component. Similar conclusions were found in other sources \citep[e.g. Ton~S180,][]{jiang18}. A more detailed comparison between low-density and high-density models for the reflection spectrum of \src\ can be found in Appendix\,\ref{low_ne}.}

\green{ \citet{caballero20} applied an intermediate  density model to both the energy spectra and the lag spectra of \src. The lamp-post geometry was used in their work. They concluded a low disc density of $10^{18}$\,cm$^{-3}$. This was because they did not include  data below 3 keV in their spectral analysis. As shown in \citet{ross07,garcia16,jiang19b}, the effects of the density parameter on the resulting reflection spectrum are, however, mostly confined to the soft X-ray band, e.g. <2\,keV.}

Despite a different density parameter, other spectral parameters are consistent with previous measurements, such as BH spin and disc inclination angle. The anti-correlation between the observed X-ray flux and the flux ratio of reflection and Comptonisation components suggests strong light-bending effects that take place in the innermost region of the accretion disc. Besides, we are also able to confirm the existence of variable UFO features that were identified in \citet{parker18}.

\red{\subsection{The Distance Between the Corona and the Disc in \src}}

\red{In this work}, we also calculate the average illumination distance between the corona and the disc by using the best-fit density and ionisation parameters. When an isotropic point-like corona model is applied, our measurements suggest that the corona moves from  \red{3$\times\frac{2\times10^{6}}{m_{\rm BH}}r_{\rm g}$} away from the disc during the lowest flux state (F1) to \red{6$\times\frac{2\times10^{6}}{m_{\rm BH}}r_{\rm g}$} during the highest flux state (F5). 

\citet{alston20} calculated the height of the corona by modelling the lag-frequency spectra extracted from the observations of \src\ in 2016 with a lamppost model: $h$ varies between $6^{+4}_{-2}$\,$r_{\rm g}$ and $15\pm3$\,$r_{\rm g}$ \red{for a similar $m_{\rm BH}\approx2\times10^{6}$}. These values of $h$ are 2--3 times higher than our estimations in Section\,\ref{size_cal}. \red{Similar measurements of $h$ were obtained by \citep{caballero20}.}

\red{Although our spectral analysis assumes no particular coronal geometry by using a broken-power law emissivity profile, the disagreement between the lamppost interpretations for the reverberation lags and the observed $n_{\rm e}\xi$ of \src\ suggests modifications to the simple lamp-post model may be needed (see Appendix\,\ref{blp} for more discussion concerning the geometry of the corona). For example, a coronal region that extends radially while moving along the rotation axis of the central black hole may exist in the NLS1 1H~0707$-$495, a sister source of \src\ \citep{wilkins14}. If such a complex coronal geometry exists in \src, the calculations of $\frac{f_{\rm AD}}{f_{\rm INF}}$ will have to change accordingly too.}

Besides, returning radiation might be another possible explanation for the difference of our measurements: some of the radiation of the disc, including thermal emission and reflection, may return to the inner disc due to strong light-bending effects. Returning radiation is particularly important around a rapidly spinning BH with $a_{*}>0.98$ \citep{cunningham76}, e.g. the supermassive BH in \src. This process will consequently cause an underestimation of $h$ by increasing the ionisation state of the inner disc. Future analysis involving multiple reflection processes is needed for the observations of \src\ and other very soft NLS1s \citep[e.g.][]{ross02}. 

\section*{Acknowledgements}

This paper was written during the COVID-19 pandemic in 2020--2021. We acknowledge the hard work of all the health care workers around the world. J.J. acknowledges supports from the Tsinghua Astrophysics Outstanding Fellowship and Tsinghua Shui'Mu Scholarship Program.

\section*{Data availability}

The data underlying this article are available in the High Energy Astrophysics Science Archive Research Center (HEASARC), at https://heasarc.gsfc.nasa.gov. \green{The \texttt{reflionx} model used in this work will be available upon reasonable request.}




\bibliographystyle{mnras}
\bibliography{iras13224.bib} 




\appendix

\section{Comparison with a low-density disc model for \src} \label{low_ne}

\red{In this section, we compare the high-density disc model in this work with a low-density disc model for the reflection component in \src.}

\red{As shown in Fig.\,\ref{pic_paras} and \citet{jiang18}, the disc reflection component of \src\ makes more contribution to the total X-ray luminosity in a low flux state than in a high flux state. Therefore, we focus on the F1 (lowest flux state) spectrum to demonstrate how the density parameter changes our model.} 

{The model considering a variable disc density parameter as introduced in Section\,\ref{mo_set} is named `Model 1' hereafter.  Model 1 is able to provide a good fit for F1 spectrum with $\chi^{2}/\nu=217.75/154$, and we find a disc density of $9\times10^{19}$\,cm$^{-3}$. The best-fit parameters are shown in the first column of Table \ref{tab_test}. They are consistent with the values obtained when we fit F1-F5 spectra simultaneously (see Table\,\ref{tab_fit}). \green{F1 spectrum alone is not able to constrain the line-of-sight Galactic column density. We thus fix this parameter at $N_{\rm H}=6.78\times10^{20}$\,cm$^{-2}$ \citep{willingale13} in the following tests.}} 

\begin{figure*}
    \centering
    \includegraphics{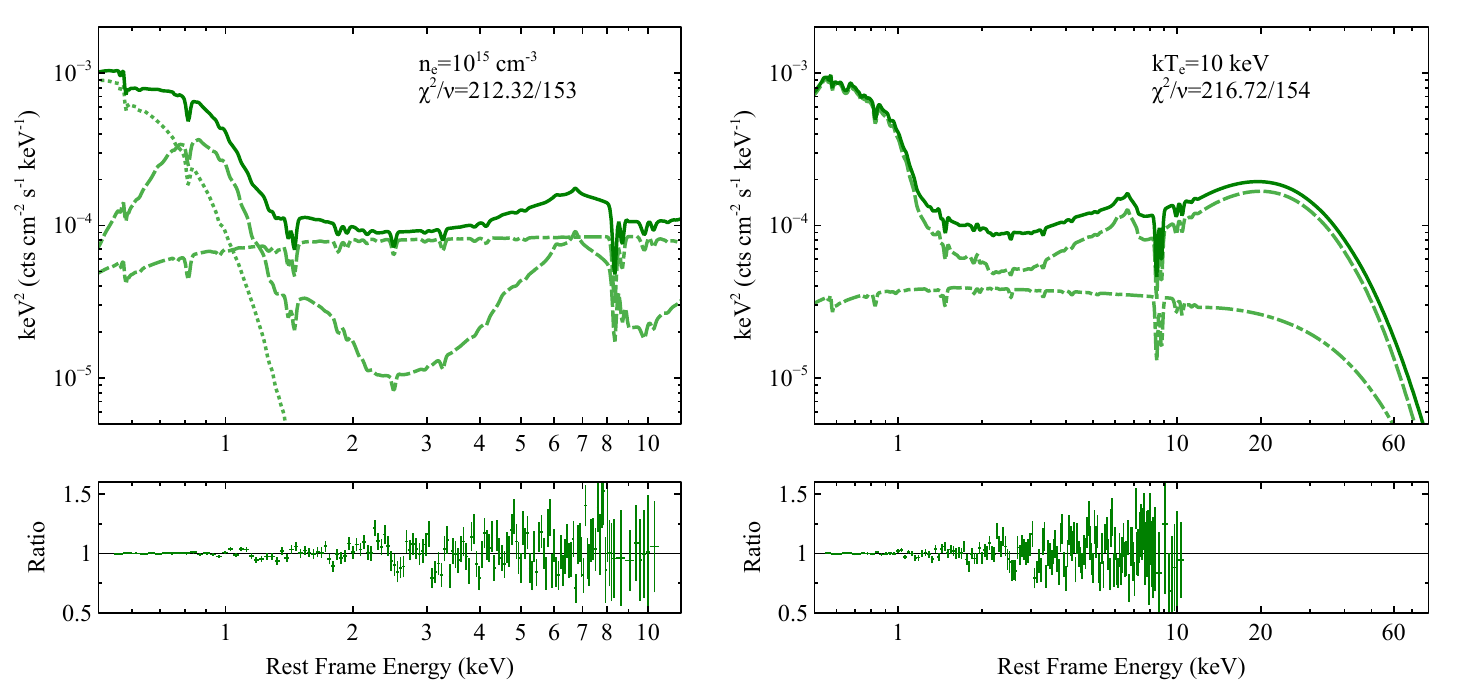}
    \caption{\red{The lowest flux state spectrum (F1) of \src\ fitted by different models. Dash-dotted lines: Comptonisation models; dashed lines: reflection models; solid lines: total models. Left: when a low electron density of $10^{15}$\,cm$^{-3}$ is used in the disc model, an additional blackbody model (\texttt{bbody}, dotted line) is required to fit the soft excess emission. \green{Right}: a cool corona of $kT_{\rm e}=10$\,keV is assumed. The disc density parameter is allowed to vary in this model. The best-fit model for the data below 10\,keV is consistent with the one based on the assumption of $kT_{\rm e}=100$\,keV.}}
    \label{pic_f0_test}
\end{figure*}

\red{We then consider a low-density disc model assuming $n_{\rm e}=10^{15}$\,cm$^{-3}$. An additional \texttt{bbody} model is required to model the soft excess emission of \src\ in combination with the relativistic disc reflection component (Model 2). The best-fit parameters are shown in the second column of Table\,\ref{tab_test}, and the best-fit model is shown in the left panel of Fig.\,\ref{pic_f0_test}. Model 2 provides a slightly better fit than Model 1 with an insignificant improvement of $\Delta\chi^{2}\approx 5$ and one additional free parameter.} \green{We only obtain an upper limit for the Galactic column density at $N_{\rm H}<9\times10^{20}$\,cm$^{-2}$ in Model 2. This parameter is fixed at $N_{\rm H}=6.78\times10^{20}$\,cm$^{-2}$ calculated by \citet{willingale13}.}

\red{In comparison, Model 2 needs a harder power-law continuum ($\Gamma\approx1.97$) than Model 1 ($\Gamma\approx2.07$). Besides, Model 2 requires a significantly higher iron abundance ($Z_{\rm Fe}$). Because the inferred disc reflection component makes less contribution to the total X-ray luminosity ($f^{\rm obs}_{\rm refl}=1.27^{+0.12}_{-0.11}$) when a low-density model is used. In comparison, $f^{\rm obs}_{\rm refl}$ for Model 1 is $3.5^{+0.4}_{-0.3}$. $Z_{\rm Fe}>20Z_{\odot}$ is thus needed in Model 2 to fit the broad Fe~K$\alpha$ emission line in the data. Similar conclusions were found in other AGNs \citep[e.g. Tons~S180,][]{jiang19b}. }

\red{Furthermore, Model 2 requires a disc ionisation parameter one order of magnitude higher than the one in Model 1. The resulting $n_{\rm e}\xi$ is only $1.6\times10^{-4}$ of the value given by Model 1. This suggests a corona that is over 90 times larger than the estimated value in Section\,\ref{calculations} (see eq.\,\ref{eq:fboostMeasured}), e.g. over a hundred $r_{\rm g}$, which is in tension with the observed reverberation lags of \src\ \citep[e.g.][]{alston20}.}

\begin{landscape}
\begin{table*}
    \centering
    \begin{tabular}{ccccccccc}
    \hline\hline
     Model & Parameters & Model 1 & \multicolumn{5}{c}{Model 2} & Model 3 \\
     \hline
     \texttt{xstar} & $N^{\prime}_{\rm H}$ ($10^{23}$\,cm$^{-2}$) & $3.2\pm0.2$ & $3.0^{+0.4}_{-0.6}$ & $3.3^{+0.6}_{-0.7}$ & $3.1^{+0.4}_{-0.5}$ & $3.0\pm0.5$ & $3.0\pm0.4$ & $3.1^{+0.3}_{-0.4}$ \\
                    & $\log(\xi^{\prime})$ (erg cm s$^{-1}$) & $3.36\pm0.03$ & $3.27^{+0.04}_{-0.08}$ & $3.37\pm0.04$ & $3.37\pm0.04$ & $3.36^{+0.03}_{-0.02}$ & $3.36\pm0.03$ & $3.32\pm0.06$\\
                    & $Z^{\prime}_{\rm Fe}$ ($Z_{\odot}$) & $3.0^{+1.0}_{-0.6}$ & $2.8^{+0.4}_{-0.5}$ & $3.2^{+1.2}_{-0.7}$ & $3.7^{+0.5}_{-1.2}$ & $3.5^{+0.4}_{-0.3}$ & $3.5\pm0.5$ & $3.0\pm0.8$ \\
                    & $z$ & $-0.158\pm0.005$ & $-0.150^{+0.006}_{-0.007}$ &  $-0.152^{+0.002}_{-0.003}$ & $-0.152\pm0.003$ & $-0.155\pm0.006$& $-0.156\pm0.004$ & $-0.153^{+0.005}_{-0.004}$ \\
     \hline
     \texttt{relconv} & q1 & $6^{+2}_{-3}$ & $>7$ & $6.5^{+1.5}_{-0.7}$ & $8.7^{+0.2}_{-1.0}$ & $>8$ & $>8$& $6\pm3$ \\
                      & q2 & $4\pm2$ & $2\pm2$ & $2.4\pm0.6$ & $2.7\pm0.4$ & $1.8^{+0.3}_{-0.2}$ & $2.1\pm0.2$ & $3.1^{+2.0}_{-1.0}$ \\
                      &$R_{\rm b}$ ($r_{\rm g}$) & $4\pm2$ & $3.2^{+1.7}_{-1.0}$ & $3\pm2$ & $3\pm2$ & $4.1^{+0.7}_{-1.2}$ & $4.8^{+0.7}_{-1.0}$ & $3.5^{+1.6}_{-0.9}$ \\
                      &$a_*$ & >0.98 & >0.98 & >0.99 & >0.99 & >0.97 & >0.97 & >0.98\\
                      &$i$ & $66\pm4$ & $67^{+2}_{-3}$ & $64\pm4$ & $69^{+3}_{-5}$ & $67^{+3}_{-4}$ & $66^{+5}_{-6}$ & $68\pm3$ \\
     \hline
     \texttt{reflionx} & $n_{\rm e}$ (cm$^{-3}$) & $9\pm2\times10^{19}$ & $10^{15}$ & \second{$10^{16}$} &\second{$10^{17}$} &\second{$10^{18}$} &\second{$10^{19}$} &$8\pm2\times10^{19}$ \\
     & $\xi$ (erg cm s$^{-1}$) & $7\pm2$ & $103^{+10}_{-8}$ & $123^{+32}_{-14}$ & $141^{+57}_{-23}$ & $76^{+45}_{-23}$ & $55^{+23}_{-17}$ & $13^{+2}_{-6}$ \\
     & $Z_{\rm Fe}$ ($Z_{\odot}$) & $3.2\pm0.5$ & >20 & >18 & $14^{+5}_{-6}$ & $8^{+4}_{-6}$ & $4.8^{+1.7}_{-0.6}$ & $3.1^{+0.7}_{-0.6}$\\
     & $\Gamma$ &  $2.072^{+0.008}_{-0.002}$ &  $1.970^{+0.009}_{-0.010}$ & $2.276^{+0.006}_{-0.005}$&$2.112^{+0.07}_{-0.008}$&$2.106^{+0.003}_{-0.004}$ & $2.067\pm0.004$ & $1.982^{+0.012}_{-0.018}$ \\
     & $\log(F_{\rm refl}$/erg\,cm$^{-2}$\,s$^{-1}$) & $-12.00^{+0.04}_{-0.03}$ & $-12.31^{+0.04}_{-0.03}$ & $-12.14\pm0.06$ & $-12.10\pm0.07$ & $-12.05\pm0.08$ & $-11.94\pm0.07$ & $-12.00^{+0.03}_{-0.02}$ \\
     & $kT_{\rm e}$ (keV) & 100 & 100 & 100 & 100& 100 & 100& 10 \\
     \hline
     \texttt{bbody} & $kT$ (eV) & - & $85^{+5}_{-2}$ & $73^{+5}_{-10}$ & $63^{+8}_{-9}$ & $62^{+4}_{-7}$ & $46\pm7$ & - \\
                    & norm ($10^{-5}$) & - & $4.6\pm0.2$ & $4.8\pm0.5$ & $7^{+4}_{-5}$ & $7\pm4$ & $3\pm2$ & -\\
     \hline
     \texttt{nthcomp} & $\log(F_{\rm pl}$/erg\,cm$^{-2}$\,s$^{-1}$) & $-12.55\pm0.03$ & $-12.412^{+0.011}_{-0.014}$ & $-12.452^{+0.010}_{-0.008}$ & $-12.38\pm0.02$ & $-12.35\pm0.03$ & $-12.46^{+0.04}_{-0.03}$ & $-12.54^{+0.02}_{-0.03}$ \\
     \hline
     & $f^{\rm obs}_{\rm refl}$ &  $3.5^{+0.4}_{-0.3}$ & $1.27^{+0.12}_{-0.11}$ & $2.0\pm0.3$ & $1.9\pm0.3$ & $2.0\pm0.4$ & $3.2^{+0.6}_{-0.5}$ & $3.5^{+0.4}_{-0.3}$ \\
     & $\chi^{2}/\nu$ & 217.75/154 & 212.32/153 & 243.66/153 & 265.71/153 & 248.63/153 & 243.66/153 & 216.72/154 \\
     \hline\hline
    \end{tabular}
    \caption{\red{Best-fit parameters for F1 spectrum of \src\ using Model 1--3. See text for more details.} \second{Model 2 considers a grid of density parameters. An additional blackbody model is used to fit the soft excess together with the disc reflection model.}}
    \label{tab_test}
\end{table*}
\end{landscape}

\begin{figure*}
    \centering
    \includegraphics[width=18cm]{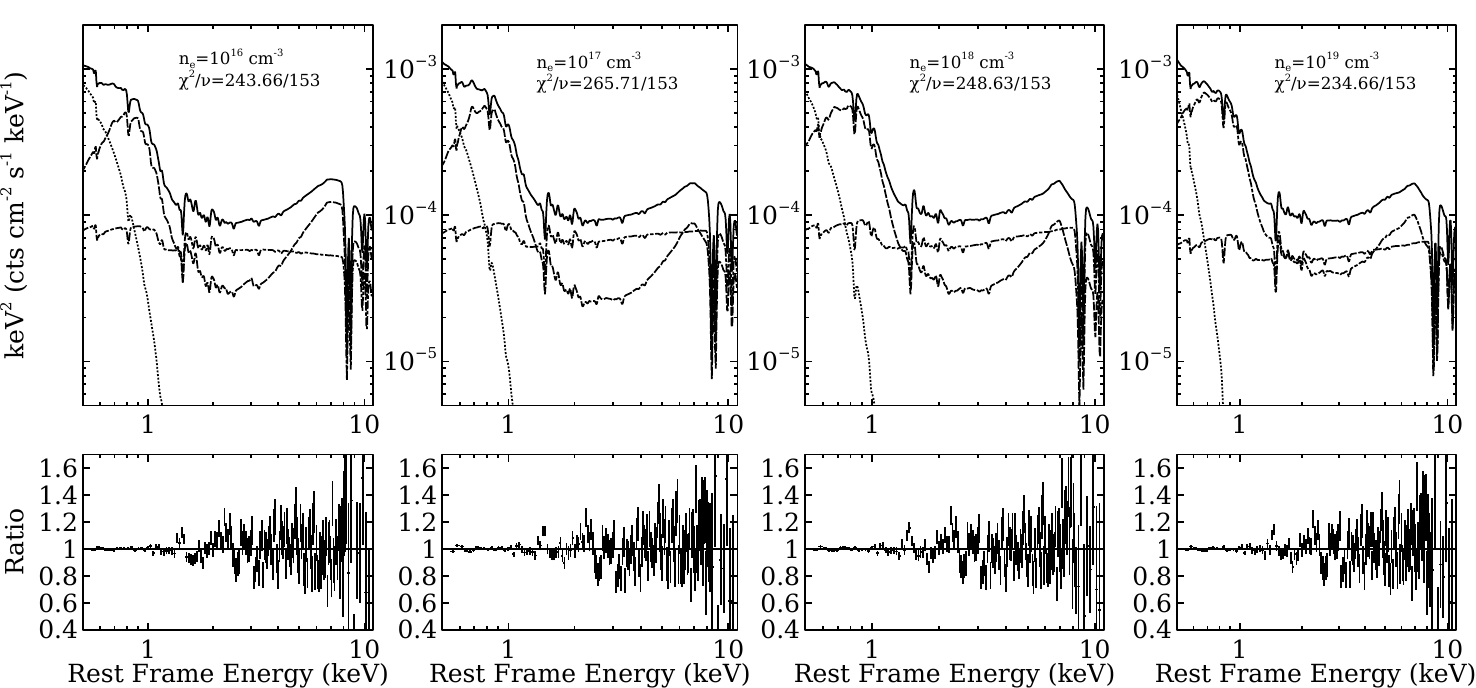}
    \caption{\second{Same as the left panel of Fig.\,\ref{pic_f0_test}. Model 2 with an intermediate density parameter. $n_{\rm e}$ is fixed at $10^{16}$\,cm$^{-3}$, $10^{17}$\,cm$^{-3}$, $10^{18}$\,cm$^{-3}$ and $10^{19}$\,cm$^{-3}$ from left to right. A zoom-in of the ratio plots for Model 2 with $10^{16}$\,cm$^{-3}$, $10^{17}$\,cm$^{-3}$ and Model 1 in the soft X-ray band is in Fig.\,\ref{pic_grid_fit_zoom}.}}
    \label{pic_grid_fit}
\end{figure*}

\begin{figure}
    \centering
    \includegraphics[width=7cm]{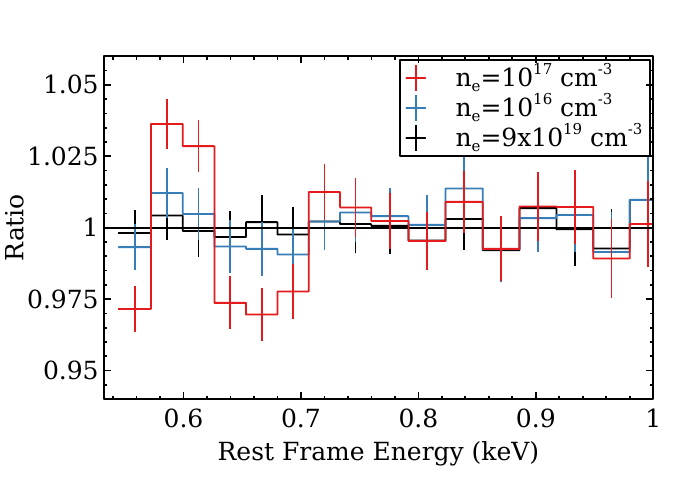}
    \caption{\second{Same as the lower panels of Fig.\,\ref{pic_grid_fit} but zoom in on the soft X-ray band for Model 2 with $n_{\rm e}=10^{16}$\,cm$^{-3}$, $10^{17}$\,cm$^{-3}$, and Model 1. A high density model improves the fit below 0.8 keV.}}
    \label{pic_grid_fit_zoom}
\end{figure}

\begin{figure*}
    \centering
    \includegraphics[width=17cm]{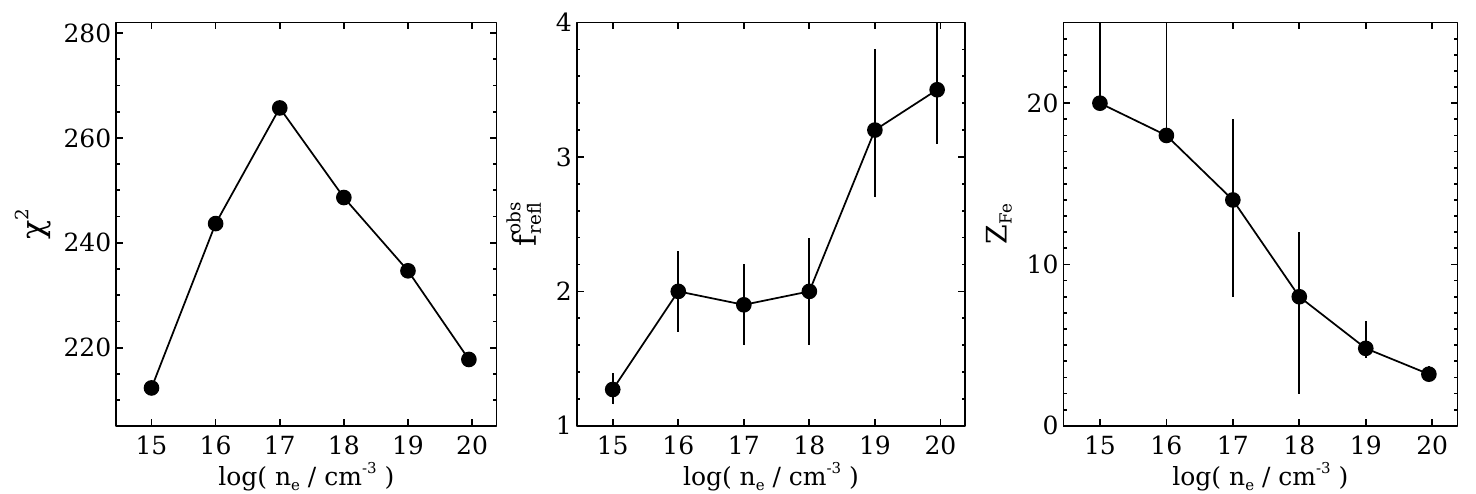}
    \caption{\second{$\chi^{2}$, $f_{\rm refl}$ and $Z_{\rm Fe}$ of the best-fit Model 2 for \src\ with a grid of density parameters. In this model, an additional \texttt{bbody} model is used to fit the soft excess emission of \src\ together with the disc reflection model. The application of high density models decreases the inferred iron abundances of the disc as the reflection component makes more contribution to the total X-ray emission of \src\ at a higher density parameter.}}
    \label{pic_grid_par}
\end{figure*}

\second{Lastly, similar to the tests in Section 4.3 in \citet{jiang18}, we investigate the possibility of fitting the spectrum of \src\ using a disc reflection model with an intermediate density parameter. Model 2, which includes an additional \texttt{bbody} model, is used for this purpose. We consider a grid of $n_{\rm e}$ by fixing this parameter at $10^{16}$, $10^{17}$, $10^{18}$ and $10^{19}$\,cm$^{-3}$.}

\second{After adapting an intermediate disc density parameter, we find the fit is significantly worse than Model 1 although they provide acceptable fits justified by $\chi^{2}/\nu$. See Table\,\ref{tab_test} for parameters and Fig.\,\ref{pic_grid_fit} for corresponding models. Model 2 with $n_{\rm e}=10^{15}$\,cm$^{-3}$ and Model 1 have $\chi^{2}=212.32$ and 217.75. In comparison, $\chi^{2}$ peaks at 265.71 when $n_{\rm e}=10^{17}$\,cm$^{-3}$. The improvement of the fits in Model 1 and Model 2 with $n_{\rm e}=10^{15}$\,cm$^{-3}$ is in the soft X-ray band, e.g. <1\,keV (see Fig.\,\ref{pic_grid_fit_zoom}). At $n_{\rm e}=10^{15}$\,cm$^{-3}$, the soft X-ray spectrum below 0.8 keV is modelled by \texttt{bbody}; at $n_{\rm e}>10^{19}$\,cm$^{-3}$, this part of the spectrum is modelled by the disc reflection component. This is because the spectral shape of the soft excess emission requires the density parameter to be within a certain range of values, e.g. $\approx9\times10^{19}$\,cm$^{-3}$ for \src, to fit the data (see Fig.\,\ref{pic_ne}). Similar conclusion was found in our previous work \citep[e.g. see Fig.\,11 in ][]{jiang18}.  When we consider the best-fit density parameter of Model 1 by fixing $n_{\rm e}$ at $9\times10^{19}$\,cm$^{-3}$ in Model 2, the \texttt{bbody} component is not constrained. We fix $kT$ of \texttt{bbody} at 46\,eV, which is the best-fit value when assuming $n_{\rm e}=10^{19}$\,cm$^{-3}$, we obtain only an upper limit of the normalisation of \texttt{bbody} ($<5\times10^{5}$).}

\second{We also note the differences in the parameters of Model 2 when different values of $n_{\rm e}$ are used. The most striking difference is the decrease of the inferred iron abundance in the reflection model along with the increase of observed reflection fraction (see Fig.\,\ref{pic_grid_par} for a comparison of these parameters). A similar result was already achieved in \citet{jiang19b}, where a significantly lower iron abundance and a higher reflection fraction were found for the NLS1 Ton~S180 when a high density model was used. The application of a high-density model increases the contribution of the reflection component indicated by the values of $f^{\rm obs}_{\rm refl}$ in the total emission. A lower iron abundance is thus required to fit the data.}

\section{The Temperature of the Corona in \src} \label{kt}

\red{During the 1.5\,Ms \xmm\ observing campaign in 2016, \nustar\ also observed \src\ with a total exposure of around 500\,ks. By fitting the 3--30\,keV FPM spectra alone with a low-density model, \citet{jiang18} found a lower limit for the high energy cutoff $E_{\rm cut}$ at 15\,keV\footnote{$E_{\rm cut}$ is approximately 2--3 times the temperature of the corona ($kT_{\rm e}$).} ($2\sigma$). We do not consider \nustar\ observations in this work, and a coronal temperature of $kT_{\rm e}=100$\,keV is used in Model 1. Because the \nustar\ observations only covered a fraction of the \xmm\ orbits and we focus on the variability of the 0.5--10\,keV emission of \src. }

\red{To investigate how a low $kT_{\rm e}$ may change our reflection model in the soft X-ray band, we fix $kT_{\rm e}$ at 10\,keV, the lowest value \texttt{reflionx} allows (Model 3). The best-fit parameters are shown in the third column of Table\,\ref{tab_test}, and the best-fit model in the 0.5--80\,keV band is shown in the middle panel of Fig.\,\ref{pic_f0_test}. When a low $kT_{\rm e}$ is used, a low exponential cut-off is seen in the model above 20\,keV. However, the fit for the soft X-ray emission is consistent except for the photon index of the continuum: a slightly lower value of $\Gamma$ ($1.982^{+0.012}_{-0.018}$) is found in comparison with a high-$kT_{\rm e}$ model ($2.072^{+0.008}_{-0.002}$).}

\section{The high density disc reflection model relxilld} \label{relxilld}

\red{Another commonly used disc reflection model is \texttt{relxilld} \citep{garcia16}, which calculates reflection spectra from a photoionised slab illuminated by a power law-shaped spectrum and includes relativistic effects in the vicinity of a BH. The allowed range of $n_{\rm e}$ in \texttt{relxilld} is $10^{15}$--$10^{19}$\,cm$^{-3}$.}

\begin{figure*}
    \centering
    \includegraphics[width=18cm]{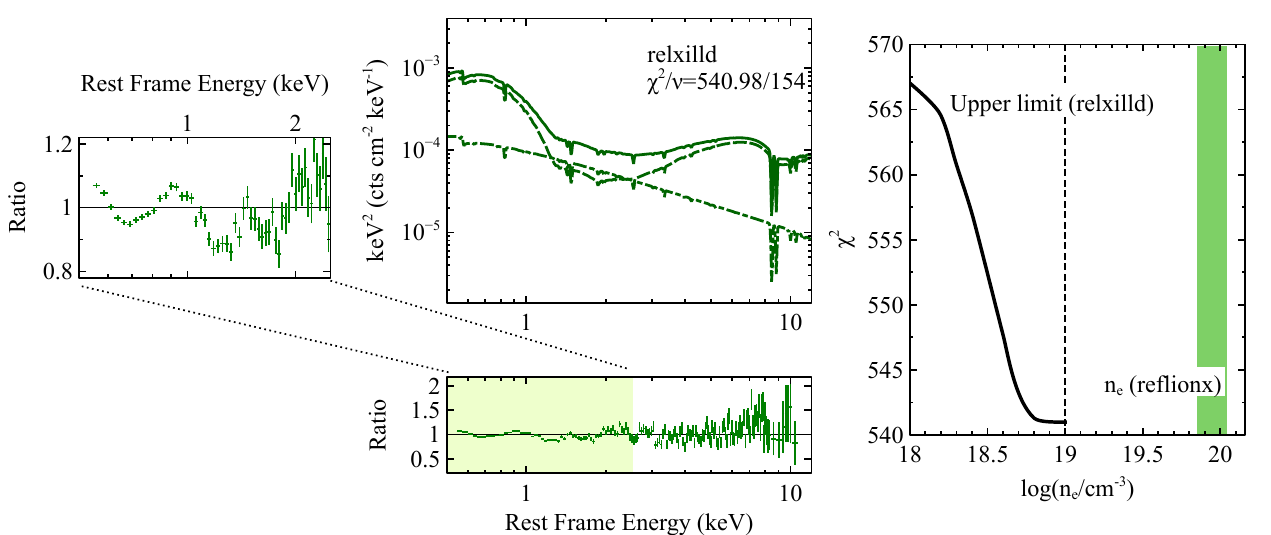}
    \caption{\red{A fit for the F1 spectrum of \src\ using the \texttt{relxilld} model. Due to the limited range of $n_{\rm e}$, the fit is pegged at $n_{\rm e}=10^{19}$\,cm$^{-3}$ as shown in the right panel, while the \texttt{reflionx} model requires $n_{\rm e}\approx10^{20}$\,cm$^{-3}$ (90\% confidence range shown by the green shaded region). Due to an inappropriate value of $n_{\rm e}$, we find it difficult to model the soft X-ray spectrum of F1 by using \texttt{relxilld}. A zoom-in on the data/model ratio plot in the soft X-ray band is shown in the left panel.}}
    \label{pic_relxilld}
\end{figure*}

\red{In Fig.\,\ref{pic_relxilld}, we present a fit for the F1 spectrum of \src\ using \texttt{relxilld}, which has the same parameters as Model 1. We find that the fit is pegged at $n_{\rm e}=10^{19}$\,cm$^{-3}$, the upper limit of the allowed parameter range in \texttt{relxilld} (see the right panel of Fig.\,\ref{pic_relxilld}).} 

\red{Due to an inappropriate value of $n_{\rm e}$, the \texttt{relxilld} model fails to reproduce the soft X-ray spectral shape in the data as shown in the ratio plot in Fig.\,\ref{pic_relxilld}. Similar conclusions were found in \citet{jiang18}: when \texttt{relxilld} was used to model the stacked spectrum of \src, the fit was pegged at $n_{\rm e}=10^{19}$\,cm$^{-3}$ and provided a significantly worse fit than other \texttt{reflionx}-based models. Therefore, we only use \texttt{reflionx} in this work.} 


\section{The Lamppost model} \label{blp}


\red{In this work, we present a disc reflection model for \src\ based on a broken-power law disc emissivity profile. No particular geometry for the corona  is assumed in our spectral model. }

\red{It has been calculated that a different coronal geometry, e.g. an aborted jet or a sphere enshrouding the inner disc, may lead to a very different disc emissivity profile and thus a different shape of the broad Fe~K$\alpha$ emission line in a disc reflection spectrum. A broken power law or a double-broken power law is a good approximation for the emissivity profiles of various geometries \citep[$F\propto r^{\rm -q}$, e.g.][]{wilkins12, gonzalez17}: most of the coronal emission reaches the innermost region of the disc due to strong light-bending effects. Then the emissivity of the disc decreases quickly with radius. At even larger radius, the emissivity profile turns flatter and approaches $r^{\rm -3}$ as in a flat, Euclidean spacetime.} 

\begin{figure*}
    \centering
    \includegraphics{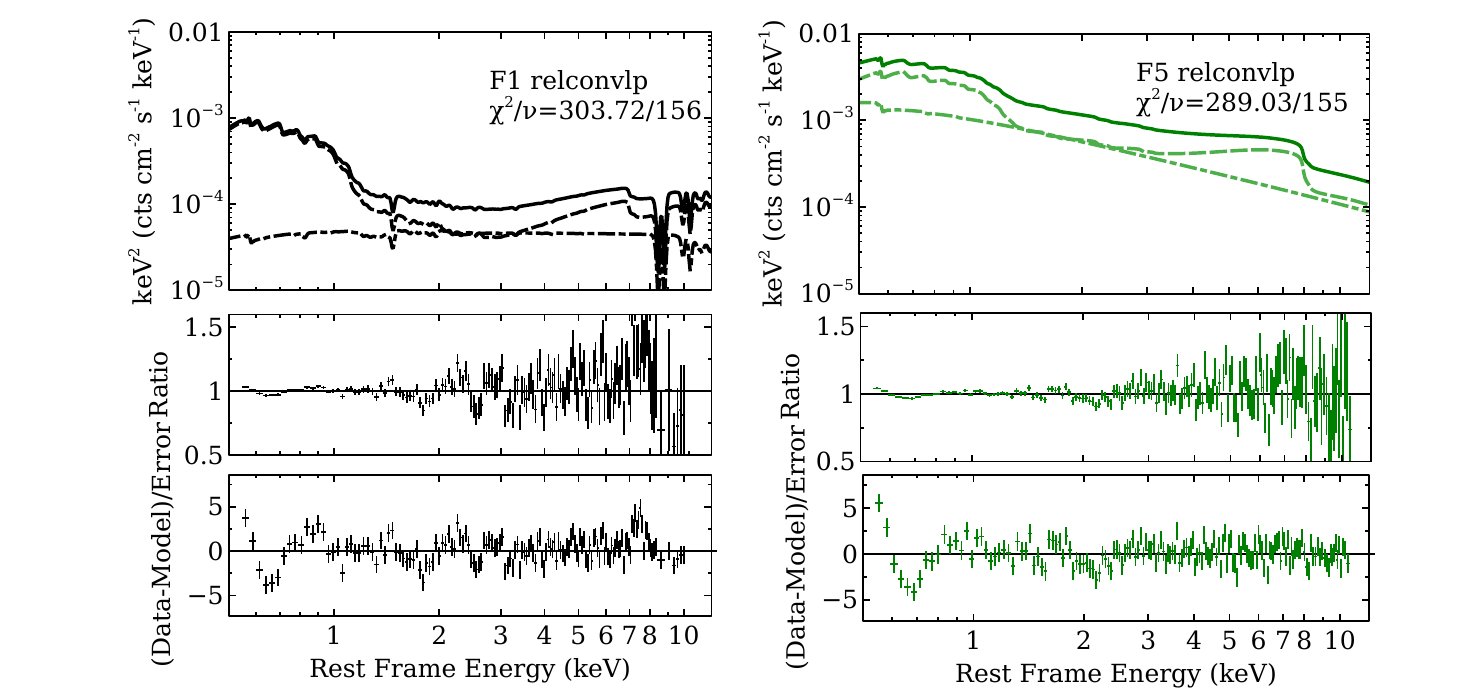}
    \caption{\green{A disc emissivity profile based on a lamppost geometry is used to fit the low-flux (F1, left) and high-flux (F5, right) spectra of \src. Residuals are seen in the soft X-ray band, e.g. <1\,keV. Solid lines: total models; dashed lines: reflection models; dash-dotted lines: Comptonisation models. Top panels show the best-fit models. The bottom two panels show corresponding data/model ratio and residual plots.}}
    \label{pic_lp_test}
\end{figure*}

{In this section, we investigate whether a lamppost model can explain the spectrum of \src. We consider \green{F1 and F5 spectra in this section to demonstrate that a simple lamp-post model is not sufficient to explain the soft X-ray spectra of \src}. 

We replace the \texttt{relconv} model in Model 1 with \texttt{relconvlp} \citep{dauser14}. The \texttt{relconvlp} model calculates the emissivity profile of a disc illuminated by an isotropic point-like source characterised by $h$, the height of the corona on the rotation axis of the BH. By applying this model to the F1 spectrum, we find a significantly worse fit with $\Delta\chi^{2}=86$ and two fewer parameters in comparison with Model 1. \green{The model is shown in the left panel of Fig.\,\ref{pic_lp_test}.} Significant residuals can be found below 1\,keV and between 7--8\,keV. We estimate the uncertainty of $h$ by using the \texttt{steppar} tool in XSPEC, and obtain $h<2.5r_{\rm g}$ ($\Delta\chi^{2}<2.706$) for F1.} \green{Similar conclusions can be found for the highest flux state spectrum where residuals can be found below 1\,keV.}



\red{In this paper, we adopt a phenomenological broken power-law emissivity as it provides a better fit to the data of \src. The disagreement between the broad band data of this source and a simple point-like corona model indicates that \green{modifications to the simple lamppost model may be required for} \src.  Similar conclusions were found in other AGNs. For example, 1H~0707$-$495, a sister source of \src, shows similar rapid X-ray variability \citep[e.g.][]{fabian09} and steep double-broken power-law disc emissivity profiles \citep{wilkins11}. Detailed modelling of its emissivity profiles indicates a coronal region that may extend in the radial direction while moving along the rotation axis of the central BH \citep{wilkins14}. Similar modelling of emissivity profiles as in \citet{wilkins14} is required to test beyond-lamppost models for \src.}

\green{The reflection fraction is also an indicator of the coronal geometry as shown by both observations \citep[e.g.][]{gallo19} and theories \citep[e.g.][]{dauser14}. A consistent calculation of the reflection fraction parameter will need to be taken into account in future beyond-lamppost spectral models for \src.} 

\green{Lastly, it is important to note that we estimate the values of $l\sqrt{f_{\rm INF}/f_{\rm AD}}$  based on the measurements of the $n_{\rm e}$ and $\xi$ parameters. This parameter increases from $0.43\times\frac{2\times10^{6}}{m_{\rm BH}}$\,$r_{\rm g}$ in the lowest flux state to $1.71\times\frac{2\times10^{6}}{m_{\rm BH}}$\,$r_{\rm g}$ in the highest flux state. These estimations are independent of the choice of the coronal geometry. In Section\,\ref{calculations2}, we adapt the over-simplified `lamp-post' model only to estimate the values of $f_{\rm INF}/f_{\rm AD}$. A different geometry may affect the values but will not change our spectral models for \src.}


\begin{table*}
    \centering
    \begin{tabular}{ccccc}
    \hline\hline
    Model & Parameter & Unit & F1 & F5 \\
    \hline
    \texttt{xstar}& $N_{\rm H}$ & $10^{23}$\,cm$^{-2}$ & $2.7^{+1.0}_{-0.3}$ & <0.2 \\
                  & $\log(\xi)$ & erg cm s$^{-1}$ & $3.41^{+0.07}_{-0.06}$ & 3.66 \\
                  & $Z^{\prime}_{\rm Fe}$ & $Z_{\odot}$ & $3.2\pm0.5$ & 2.7 \\
                  & $z$ & - & $-0.147\pm0.006$ & -0.165 \\
    \hline
    \texttt{relconvlp} & h$^{1}$ & $r_{\rm g}$ & \second{1.9 (<3)} & \second{3.2 (<6)} \\
                       & $i$ & deg & $62^{+3}_{-2}$ &  $57^{+8}_{-10}$\\
                       & $a_{*}$ & - & >0.98 & >0.95 \\
    \hline
    \texttt{reflionx} & $n_{\rm e}$ & 10$^{20}$ cm$^{-3}$ & $1.0^{+0.2}_{-0.3}$ & $1.2^{+0.3}_{-0.4}$ \\
                      & $\xi$ & erg cm s$^{-1}$ & $7\pm2$ & $12^{+5}_{-6}$ \\
                      & $Z_{\rm Fe}$ & $Z_{\odot}$  & $3.5^{+0.6}_{-0.7}$ & $2.8^{+0.7}_{-0.2}$ \\
                      & $\Gamma$ & - & $2.09^{+0.02}_{-0.03}$ & $2.99^{+0.02}_{-0.03}$ \\
                      & $\log(F_{\rm refl})$ & erg\,cm$^{-2}$\,s$^{-1}$ & $-11.937^{+0.010}_{-0.012}$ & $-11.23\pm0.03$ \\
    \hline
    \texttt{nthcomp} & $\log(F_{\rm pl})$ & erg\,cm$^{-2}$\,s$^{-1}$ & \second{$-12.64^{+0.06}_{-0.03}$} & $-11.50\pm0.02$ \\
    \hline
    & $f^{\rm obs}_{\rm refl}$ & - & \second{$5.0^{+0.6}_{-0.4}$} & $1.9^{+0.5}_{-0.4}$ \\
    & $\chi^{2}/\nu$ & - & 303.72/156 & 289.03/155 \\
    \hline\hline
    \end{tabular}
    \caption{\green{Best-fit parameters for the low-flux (F1) and high-flux (F5) spectra of \src\ assuming a lamp-post emissivity profile. This model provides worse fits to the data than a model using broken power-law emissivity profiles. \second{$^{1}$ Only upper limits are obtained for this parameter and shown in the brackets.}}}
    \label{tab_lp}
\end{table*}

\section{A DOUBLE-REFLECTION, DOUBLE-ABSORPTION MODEL} \label{stack}

\red{\src\ shows rapid variability in the X-ray band on timescales of kiloseconds \citep[e.g.][]{alston20}. During the \xmm\ observing campaign in 2016, the observed EPIC-pn count rate changed by up to 2 orders of magnitude on timescales of days \citep{jiang18}, making it one of the most extreme known AGNs.}

\red{To study its spectral variability, we divide the observations of \src\ into five flux intervals. 
In Section\,\ref{mo_set}, we introduce a disc reflection model with an electron density of around $10^{20}$\,cm$^{-3}$ to explain each flux-resolved spectrum of \src. Alternatively, one may find an equally good fit using a low-density model as introduced in Section\,\ref{low_ne}. However, an additional \texttt{bbody} model is required to fit the soft excess emission when $n_{\rm e}=10^{15}$\,cm$^{-3}$ is assumed. Either a high-density or a low-density model has only one reflection component. Besides, one photoionisation model \texttt{xstar} is needed to model the blueshifted Fe~\textsc{xxv}/\textsc{xxvi} absorptions in the 8--9\,keV band of flux-resolved spectra.}

\red{In this section, we argue that the requirement for a double-reflection, double-absorption model as in \citet{chiang15, jiang18} is due to an averaged spectrum extracted from a wide range of flux states being used in the previous work. To test this idea, we consider a spectrum from F1 and F4\footnote{F1 represents the lowest flux interval, and F4 represents the second highest flux interval. We do not use F5 interval in this test because F5 does not show significant UFO absorptions as shown in Section\,\ref{data}.} intervals combined.}

\begin{figure*}
    \centering
    \includegraphics[width=18cm]{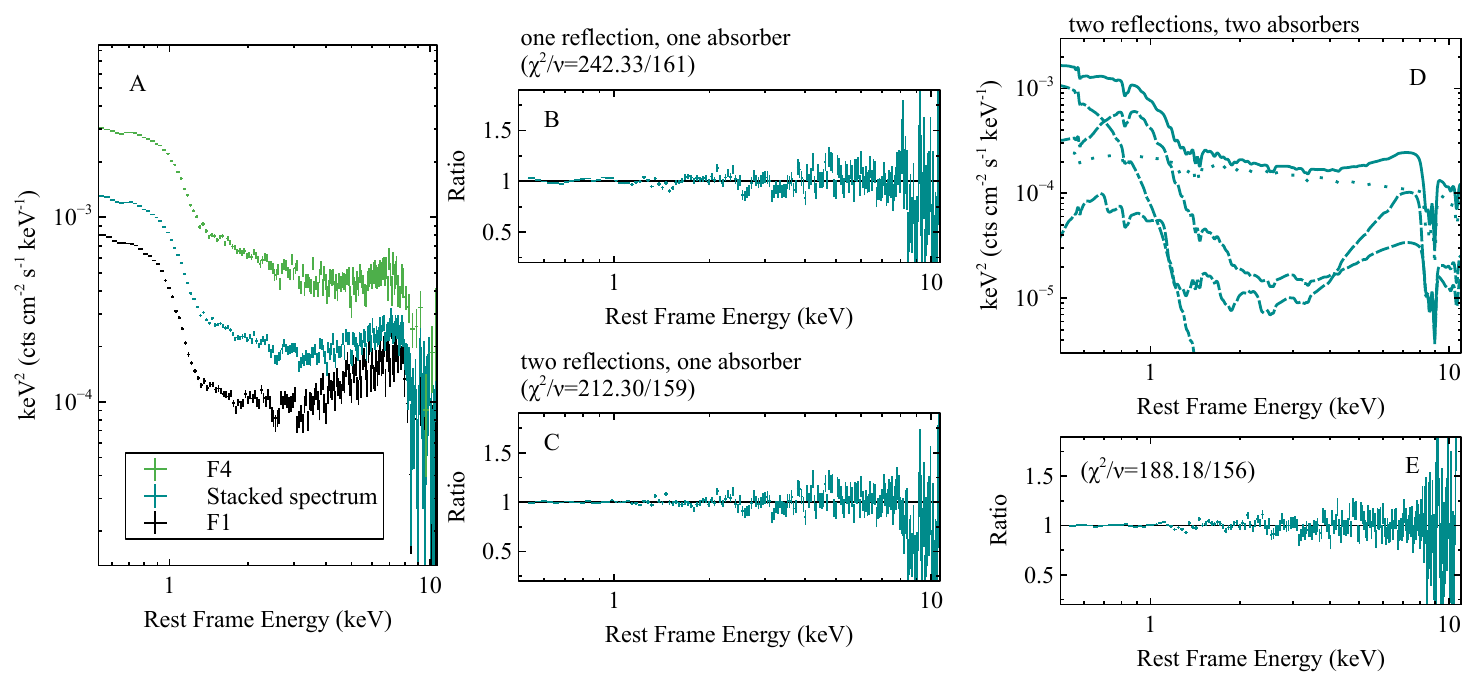}
    \caption{\red{Panel A: unfolded spectra of \src\ extracted from F4 (top), F1+F4 (middle) and F1 (bottom) intervals. The spectra are unfolded using a model that is constant across the energy band for demonstration purposes only. Panel B: a data/model ratio plot for the F1+F4 spectrum using a single-reflection, single-absorption model. Significant positive residuals are seen in the 7--8\,keV band and negative residuals are seen at 9\,keV. Panel C: a data/model ratio plot for the same stacked spectrum using a double-reflection, single-absorption model. Negative residuals are still found at 9\,keV. Panel D: the best-fit double-reflection, double-absorption model for the stacked spectrum of \src. Corresponding data/model ratio plot is shown in the panel E. This model is similar to the ones in \citet{jiang18}.}}
    \label{pic_stack}
\end{figure*}

\red{The stacked F1+F4 spectrum of \src\ is shown in the Panel A of Fig.\,\ref{pic_stack} in comparison with F1 and F4 spectra. We follow the same analysing approach as in \citet{jiang18} by starting with a single-reflection, single-absorption model. An electron density of $10^{15}$\,cm$^{-3}$ is used as in \citet{chiang15} and \citet{jiang18}. The full model is \texttt{tbnew * xstar * ( relconv * reflionx + bbody + nthcomp)} in XSPEC notation. The line-of-signt Galactic column density is fixed at $N_{\rm H}=6.78\times10^{20}$\,cm$^{-2}$ \citep{willingale13} in this test. We find that this single-reflection, single-absorption model fails to reproduce the observed blue wing of the broad Fe~K$\alpha$ emission and the absorption features above 7\,keV (see the Panel B of Fig.\,\ref{pic_stack}). An additional reflection component is able to significantly improve the fit by $\Delta\chi^{2}=30$ and two more free parameters, which are the ionisation and the normalisation parameters of the second reflection component. The other parameters, including $Z_{\rm Fe}$, emissivity indices, $a_{*}$ and $i$ are linked to the corresponding parameters of the first reflection component. However, a second absorption model is still required to fit the rest of the broad absorption feature above 8.5\,keV. The width of the absorption line is due to the averaging effects of rapidly variable UFOs as seen in \citet{parker17b}.}

\red{In the end, we present the best-fit model for the stacked spectrum \src\ by considering the same model set-up as in \citet{jiang18}. A double-reflection, double-absorption model is required and able to provide a good fit to the stacked spectrum with $\chi^{2}/\nu=188.18/156$. The best-fit parameters are shown in Table\,\ref{tab_stack}, and the best-fit model is shown in the Panel D of Fig.\,\ref{pic_stack}. Corresponding data/model ratio plot can be found in the Panel E of Fig.\,\ref{pic_stack}.} 

\red{So far, we have successfully reproduced the spectral fitting process in \citet{jiang18} by considering a stacked spectrum of \src\ extracted from low and high flux intervals combined. Due to the averaging effects, a double-reflection, double-absorption model is needed. We conclude that, in order to study the spectral variability of a variable AGN like \src, one needs to avoid stacking spectra from a wide range of flux intervals.}

\begin{table}
    \centering
    \begin{tabular}{ccccc}
    \hline\hline
     Model & Parameters & Values \\
     \hline
     \texttt{xstar1} & $N^{\prime}_{\rm H}$ ($10^{23}$\,cm$^{-2}$) & $7.0\pm0.2$  \\
                    & $\log(\xi^{\prime})$ (erg cm s$^{-1}$) &  $3.70^{+0.08}_{-0.09}$ \\
                    & $Z^{\prime}_{\rm Fe}$ ($Z_{\odot}$) & $2.9^{+0.3}_{1.4}$  \\
                    & $z$ & $-0.178^{+0.006}_{-0.008}$ \\
        \hline
      \texttt{xstar2} & $N^{\prime}_{\rm H}$ ($10^{23}$\,cm$^{-2}$) & $4.6^{+2.1}_{-0.3}$  \\
                    & $\log(\xi^{\prime})$ (erg cm s$^{-1}$) &  $3.48^{+0.07}_{-0.06}$ \\
                    & $z$ & $-0.148^{+0.005}_{-0.003}$ \\
     \hline
     \texttt{relconv} & q1 & >6 \\
                      & q2 &  $2.1^{+1.4}_{-0.6}$ \\
                      &$R_{\rm b}$ ($r_{\rm g}$) &  $2.0^{+1.2}_{-0.4}$ \\
                      &$a_*$ & >0.98 \\
                      &$i$ & $64^{+2}_{-3}$  \\
     \hline
     \texttt{reflionx1} & $\xi$ (erg cm s$^{-1}$) & $25^{+12}_{-7}$\\
     & $Z_{\rm Fe}$ ($Z_{\odot}$) & >18 \\
     & $\Gamma$ &  $2.41\pm0.04$ \\
     & $\log(F_{\rm refl}$/erg\,cm$^{-2}$\,s$^{-1}$) & $-12.71^{+0.09}_{-0.08}$ \\
    \hline
     \texttt{reflionx2} & $\xi$ (erg cm s$^{-1}$) & $226^{+30}_{-12}$\\
     & $\log(F_{\rm refl}$/erg\,cm$^{-2}$\,s$^{-1}$) & $-12.12^{+0.06}_{-0.09}$ \\
     \hline
     \texttt{bbody} & $kT$ (eV) & $84^{+5}_{-6}$ \\
                    & norm ($10^{-5}$) &  $5\pm0.2$ \\
     \hline
     \texttt{nthcomp} & $\log(F_{\rm pl}$/erg\,cm$^{-2}$\,s$^{-1}$) & $-12.07\pm0.02$ \\
     \hline
     & $\chi^{2}/\nu$ & 188.18/156 \\
     \hline\hline
    \end{tabular}
    \caption{\red{Best-fit parameters for F1+F4 spectrum of \src\ using the same model in \citet{jiang18}. The full model is \texttt{tbnew * xstar1 * xstar2 * ( relconv * (reflionx1 + reflionx2) + bbody + nthcomp)}.}}
    \label{tab_stack}
\end{table}

\bsp	
\label{lastpage}
\end{document}